\documentclass{ws-ijmpe}
\usepackage[super,compress]{cite}
  \usepackage{amssymb}
   \usepackage{amsmath}
    \usepackage{amsfonts}
     \usepackage{epsfig}
     \usepackage{subfigure}
      \usepackage{bm} 

 \def\gsim{\mathrel{\rlap{\lower4pt\hbox{\hskip1pt$\sim$}}
 \raise1pt\hbox{$>$}}}

 \renewcommand{\floatsep} {-3cm}

 \newcommand\ra{\rangle}
 \newcommand\beq{\begin{equation}}
 
 \newcommand\eeq{\end{equation}}
 \newcommand\beqn{\begin{eqnarray}}
 \newcommand\eeqn{\end{eqnarray}}

\def\mb{\,\mbox{mb}}
\def\fm{\,\mbox{fm}}
\def\GeV{\,\mbox{GeV}}

\def\lsim{\mathrel{\rlap{\lower4pt\hbox{\hskip1pt$\sim$}}
    \raise1pt\hbox{$<$}}}         
\def\gsim{\mathrel{\rlap{\lower4pt\hbox{\hskip1pt$\sim$}}
    \raise1pt\hbox{$>$}}}         

\def\mb{\,\mbox{mb}}
\def\fm{\,\mbox{fm}}
\def\GeV{\,\mbox{GeV}}

\def\beq{\begin{equation}}
\def\eeq{\end{equation}}

\def\beqy{\begin{eqnarray}}
\def\eeqy{\end{eqnarray}}

\begin{document}

\markboth{Boris Kopeliovich, Roman Pasechnik, Irina Potashnikova}{Hard hadronic diffraction is not hard}

\catchline{}{}{}{}{}

\title{Hard hadronic diffraction is not hard}

\author{Boris Kopeliovich}

\address{Departamento de F\'{\i}sica,
Universidad T\'ecnica Federico Santa Mar\'{\i}a;\\
Centro Cient\'ifico-Tecnol\'ogico de Valpara\'{\i}so,
Casilla 110-V, Valpara\'{\i}so, Chile \\
bzk@mpi-hd.mpg.de}

\author{Roman Pasechnik}

\address{Department of Astronomy and Theoretical Physics, Lund
University, SE-223 62 Lund, Sweden\\
Roman.Pasechnik@thep.lu.se}

\author{Irina Potashnikova}

\address{Departamento de F\'{\i}sica,
Universidad T\'ecnica Federico Santa Mar\'{\i}a;\\
Centro Cient\'ifico-Tecnol\'ogico de Valpara\'{\i}so,
Casilla 110-V, Valpara\'{\i}so, Chile \\
irina.potashnikova@usm.cl}

\maketitle

\begin{history}
\received{Day Month Year}
\revised{Day Month Year}
\end{history}

\begin{abstract}
Hadronic diffractive processes characterised by a hard scale (hard diffraction) contain a nontrivial interplay of hard and soft, nonperturbative interactions, 
which breaks down factorisation of short and long distances. On the contrary to the expectations based on the factorization hypothesis, assuming that hard 
diffraction is a higher twist, these processes should be classified as a leading twist. We overview various implications of this important 
observation for diffractive radiation of Abelian (Drell-Yan, gauge bosons, Higgs boson) and non-Abelian (heavy flavors) particles, as well as direct 
coalescence into the Higgs boson of the non-perturbative intrinsic heavy flavour component of the hadronic wave function.
\end{abstract}


\ccode{PACS numbers: 13.87.Ce, 14.65.Dw, 14.80.Bn}

\section{Introduction}
\label{Sec:Intro}

Nowadays, factorisation of hard and soft interactions in Quantum Chromo Dynamics (QCD) in inclusive reactions is one of the most powerful
and widely used theoretical tools [\refcite{factorization}]. While soft long-range interactions are poorly known, they factorise from well-studied
short-distance interactions. Typcally, the soft part is represented in terms of parton distribution functions (PDFs) whose dependence
on the hard factorisation scale is provided by evolution equations with non-perturbative (supposedly, universal) starting distributions 
parameterised from the data. Making a plausible (although, not rigorously proven) assumption about universality of the soft interactions, 
one often constrains those with electro-weak hard probes such as the Deep Inelastic Scattering (DIS) and Drell-Yan (DY) processes and then 
applies to hard hadronic processes. One may naturally be tempted to extend such a factorization scheme to diffractive reactions with a large 
rapidity gap. However, such diffractive factorisation turns out to be unavoidably broken and does not hold in practice [\refcite{rev1,rev2}].

Inelastic diffraction in the Good-Walker picture is  treated as a shadow of inelastic processes [\refcite{Glauber,FP56,GW}]. Consider
the incoming plane wave which scatters off the target. If it has differently interacting components, the outgoing wave will necessarily
have a different composition, namely, a new (diffractive) state emerges (for more details, see Ref.~[\refcite{rev1,kaidalov-rev}]). 
Purely soft hadronic diffraction incorporates unknown non-perturbative physics and thus is theoretically very difficult to predict.
Instead, diffractive processes involving a hard scale attract particular attention. Although factorisation of interactions at short 
and long distances still holds in diffractive DIS, the corresponding fracture functions are proven not be universal; they are 
process-dependent and hence cannot be used for other diffractive processes.

\begin{figure*}[h]
\centering
\includegraphics[width=3.5cm,clip]{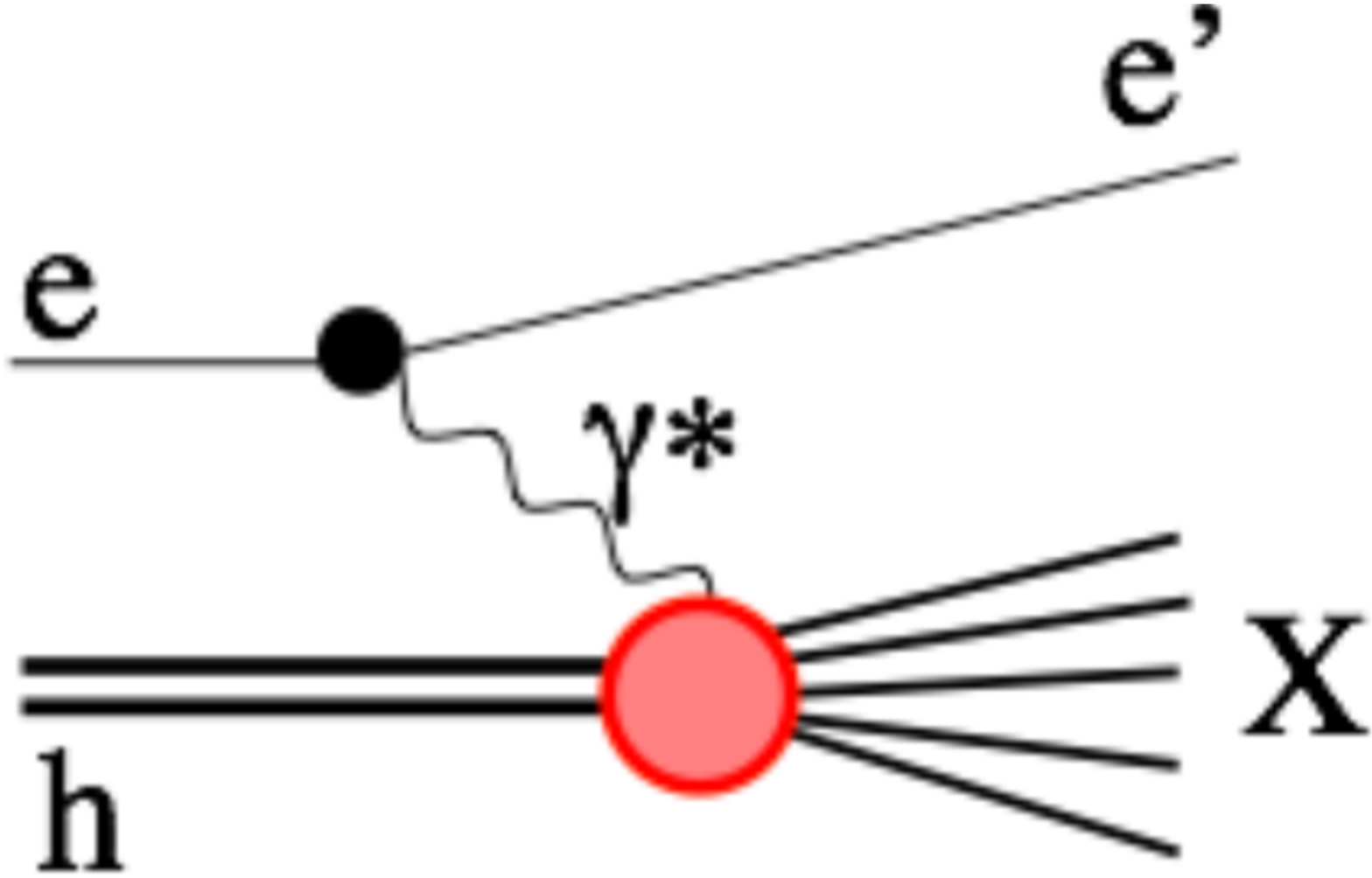}
  \hspace*{2cm}
\includegraphics[width=4cm,clip]{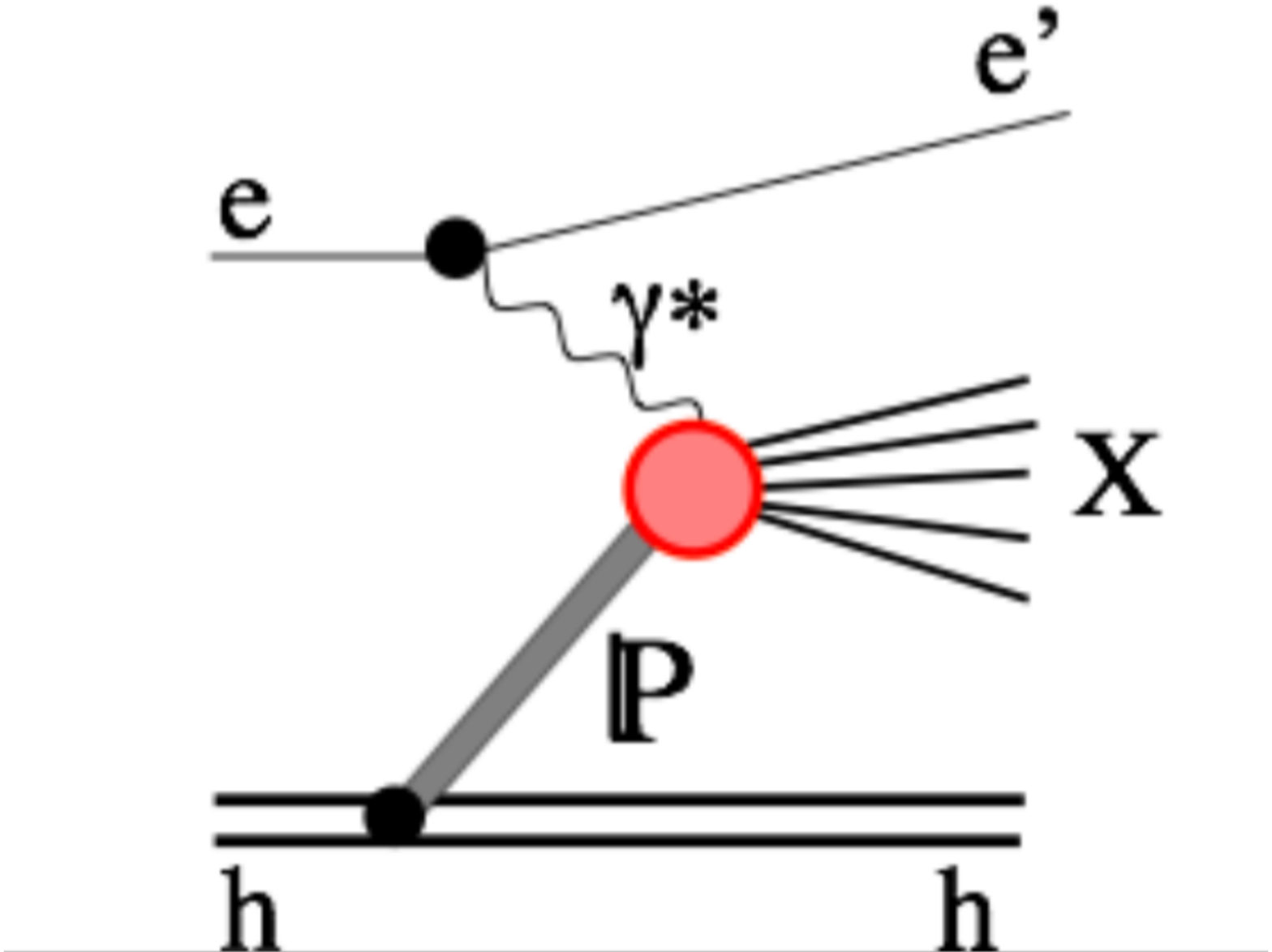}
\caption{An illustration of the DIS process on hadronic taget (left panel) 
and on the Pomeron which is often treated as a target (right panel).}
\label{fig:dis}       
\end{figure*}

Traditionally, the Pomeron-hadron total cross section is studied by means of diffractive excitations [\refcite{kaidalov-rev}]. Being applied
to DIS, it enables to constrain the structure function of the Pomeron directly from the data [\refcite{is}]. One way to study diffractive processes
is to extrapolate the ideas of QCD factorisation by employing the PDFs in the Pomeron which would, in principle, allow to predict the hard
diffractive cross sections in hadronic collisions as is illustrated in Fig.~\ref{fig:dis}. It is well-known, however, that such predictions for
hard diffractive observables, for example, on high-$p_T$ dijet production, strongly fail by about an order of magnitude (see e.g. Refs.~
[\refcite{tev-1,tev-2}]). This happens, in particular, due to a strong breakdown of diffractive factorisation due to the unavoidable presence 
of spectator partons at typically large (hadronic) separations [\refcite{KPST-06}].

Within the quantum-mechanical Good-Walker mechanism of diffraction [\refcite{Glauber,FP56,GW}], the off-diagonal amplitude should 
be treated as a linear combination of diagonal (elastic) diffractive amplitudes corresponding to different Fock components in 
the projectile hadronic wave function. In this picture, a result of cancellations between various elastic amplitudes,
the diffractive amplitude exhibits a common factor giving rise to the absorptive corrections which, thus, naturally
emerge at the amplitude level, and do not require an additional gap survival factors in the cross section. This is
especially transparent and clear in the color dipole approach [\refcite{zkl,nik}] where any diffractive scattering is viewed
as a superposition of elastic scatterings of ${\bar q}q$ dipoles of varied sizes arising from a combination of
projectile Fock states in the initial hadronic wave function. As was explicitly demonstrated for the first time 
in Ref.~[\refcite{KPST-06}], the long-range soft interactions with spectators comes together with hard interactions
in the DY reactions associated with a short-distance virtual $\gamma^*$-radiation, and the latter two cannot be
consistently separated.

In this way, any diffractive amplitude emerges as a difference between the elastic scatterings of different Fock
components in the projectile composite state, in particular, that of hadronic states with and without 
a hard fluctuation (for example, $qq$ and $qq\gamma^*$ states in the DY case),
\beq
 A_{\rm diff}^{\rm DY}\propto \sigma_{\bar qq}(\vec R+\vec r)-\sigma_{\bar qq}(\vec R) 
 \propto \vec r\cdot \vec R\sim 1/Q \,, \label{40}
\eeq
where $\sigma_{\bar qq}(r)$ is the universal dipole-nucleon cross section [\refcite{zkl,nik}] fitted to DIS data, $\vec R$ corresponds to a large separation 
between different constituent projectile quarks in the incoming hadron while small $r\sim 1/Q\ll R$ is related to the hard radiation 
process (see e.g. Refs.~[\refcite{dy-k,dy-r}]). Obviously, such a mild hard scale $Q$ dependence, corresponding to the leading-twist behavior, 
strongly contradicts to the diffractive factorisation DIS-like prediction,
\beq
A_{\rm diff}^{\rm DIS} \propto \sigma_{\bar qq}(r) \propto r^2\sim 1/Q^2 \,, \label{20}
\eeq
which is apparently a higher-twist effect. While the phenomenological dipole cross section (or partial dipole amplitude) is a universal 
ingredient naturally accumulating the soft interactions and fitted to the available precision data, the diffractive amplitudes are represented 
in terms of a linear superposition of elastic dipole scatterings at different transverse separations which is process-dependent and accumulates 
all the relevant absorptive corrections fully dynamically. Naturally, the gap survival amplitude gets singled out from such a superposition as 
a common factor dependent on soft parameters of the dipole model and on the diffractive process concerned.

In the forward scattering limit and in the absence of spectator co-movers a single quark cannot radiate an Abelian particle ($\gamma,\,Z,\,W^\pm,\,H$) 
in a diffractive quark-hadron scattering (with zero net momentum transfer), in variance to diffractive factorisation [\refcite{KST-par}]. Only a dipole can 
diffractively radiate due to a small fluctuation in its size induced by the hard scattering (c.f. Eq.~(\ref{40})) so diffraction becomes possible
although is strongly suppressed. Such a mechanism opens up new possibilities for universal description of diffractive reactions specific to the
dipole approach beyond QCD factorisation [\refcite{zkl}]. The diffractive factorisation breaking in non-Abelian radiation is also important
although the diffractive gluon radiation off a quark does not vanish in the forward kinematics due to an extra glue-glue interaction. 
The universal dipole mechanism of diffraction has been employed in a number of diffractive processes so far, and this review aims at 
a short comprehensive overview of major implications of the diffractive factorisation breakdown in both Abelian and non-Abelian 
diffractive radiation.

\section{Color dipole picture}
\label{Sec:dip}

The color dipole formalism [\refcite{zkl,al,nik}] is a phenomenological approach which effectively takes into account 
the higher-order QCD corrections in dipole-target scattering in the target rest frame. Color dipoles with definite transverse 
sizes are treated as eigenstates of interaction, thus, their scattering off a target is universal and can be used as a building block
for more complicated inclusive and diffractive processes. The dipole cross section is the basic ingredient of this
approach which is process-independent and is typically extracted from ample DIS phenomenology. In the simplest 
form, without an account for QCD evolution, the dipole cross section at small Bjorken variable $x<0.01$ is conventionally 
parameterized in the following saturated form known as the Golec-Biernat--Wusthoff (GBW) ansatz [\refcite{GBW}],
\begin{eqnarray}
 \sigma_{\bar qq}(\vec r,x) = \int d^2b\,2{\rm Im} f_{\rm el}(\vec b,\vec r) = \sigma_0\, (1-e^{-r^2/R_0^2(x)})\,,
 \label{Sig-dip}
\end{eqnarray}
whose parameters $\sigma_0=23.03\mb$ and $R_0(x)=0.4\fm\times(x/x_0)^{0.144}$, $x_0=0.003$ 
are known from fits to the DIS data, and $f_{\rm el}(\vec b,\vec r)$ is the partial dipole amplitude. Such a simplified 
parametrisation (cf. Ref.~[\refcite{bartels}]) provides a reasonable description of all bulk of inclusive and diffractive DIS data 
in $ep$ collisions at HERA as well as many other processes in hadron-hadron and hadron-nucleus collisions such as DY, 
heavy quark and prompt photon production etc [\refcite{nik,k95,bhq97,kst99,krt01,npz}]. 

The dipole cross section (\ref{Sig-dip}) levels off at $r\gg R_0$, the phenomenon commonly known as saturation.
The second important feature is that it vanishes at small $r\to 0$ as $\sigma_{q\bar q}\propto r^2$ [\refcite{zkl}]
gising rise to the color transparency propery. The latte reflects the fact that a point-like colorless object does not 
interact with external color fields. Finally, the quadratic $r$-dependence at small $r$ is a natural consequence of 
gauge invariance and nonabeliance of interactions in QCD.

In soft diffractive processes, the ansatz (\ref{Sig-dip}) has to be modified since the Bjorken variable $x$ is not
a proper variable in this case. Instead, the gluon-target collision c.m. energy squared $\hat s=x_1 s$ given in 
terms of the gluon momentum fraction $x_1$ and the $pp$ c.m. energy $s$) is a more appropriate variable, while 
the saturated form (\ref{Sig-dip}) is retained. An analytic form of the $x$- and $\hat s$-dependent 
parameterisations for the partial amplitude $f_{\rm el}(\vec{b},\vec{r})$ can be found e.g. in 
Refs.~[\refcite{GBW-par,kpss,KST-GBW-eqs}].

\section{Diffractive Abelian radiation}
\label{Sec:dy}

\begin{figure}[t]
\centering
\subfigure[]{
\scalebox{0.24}{\includegraphics{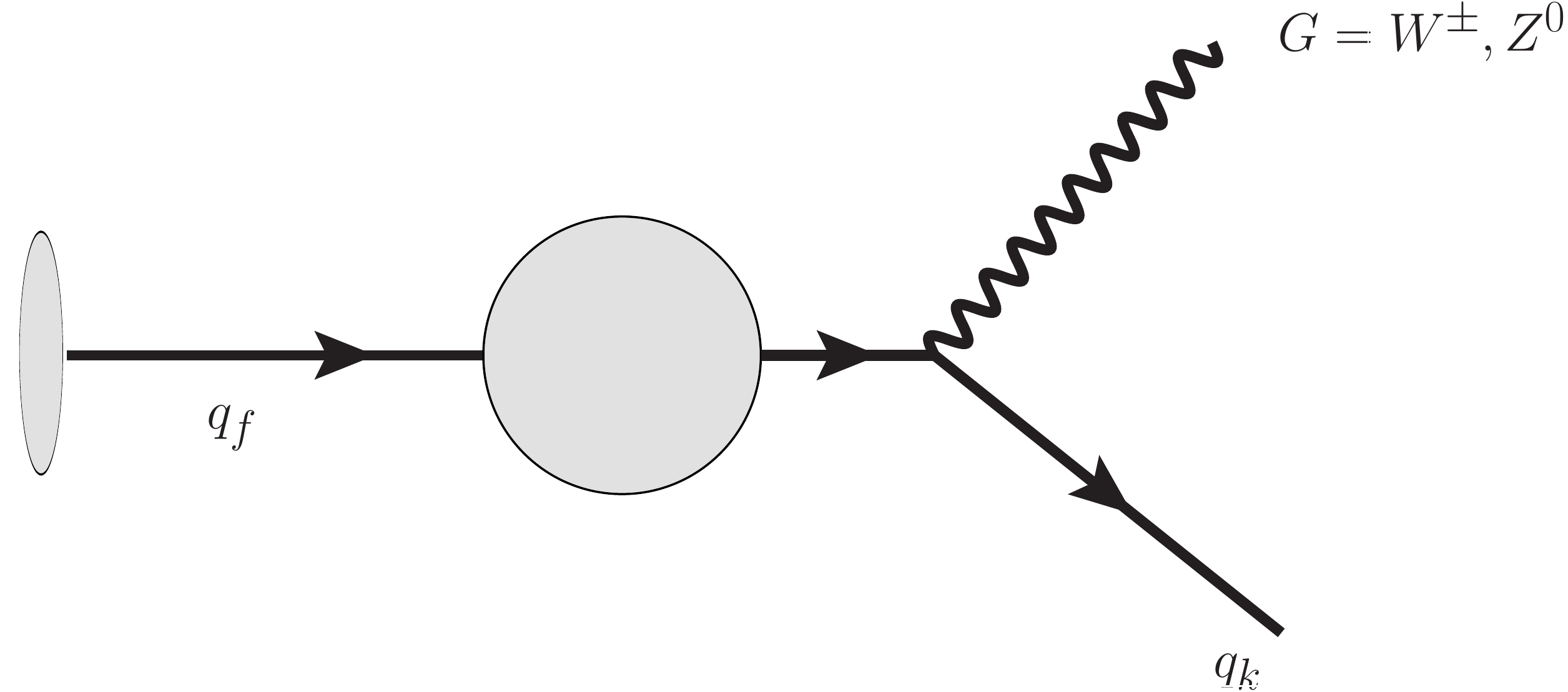}}
\label{fig:gb_dir}
}
\centering
\subfigure[]{
\scalebox{0.24}{\includegraphics{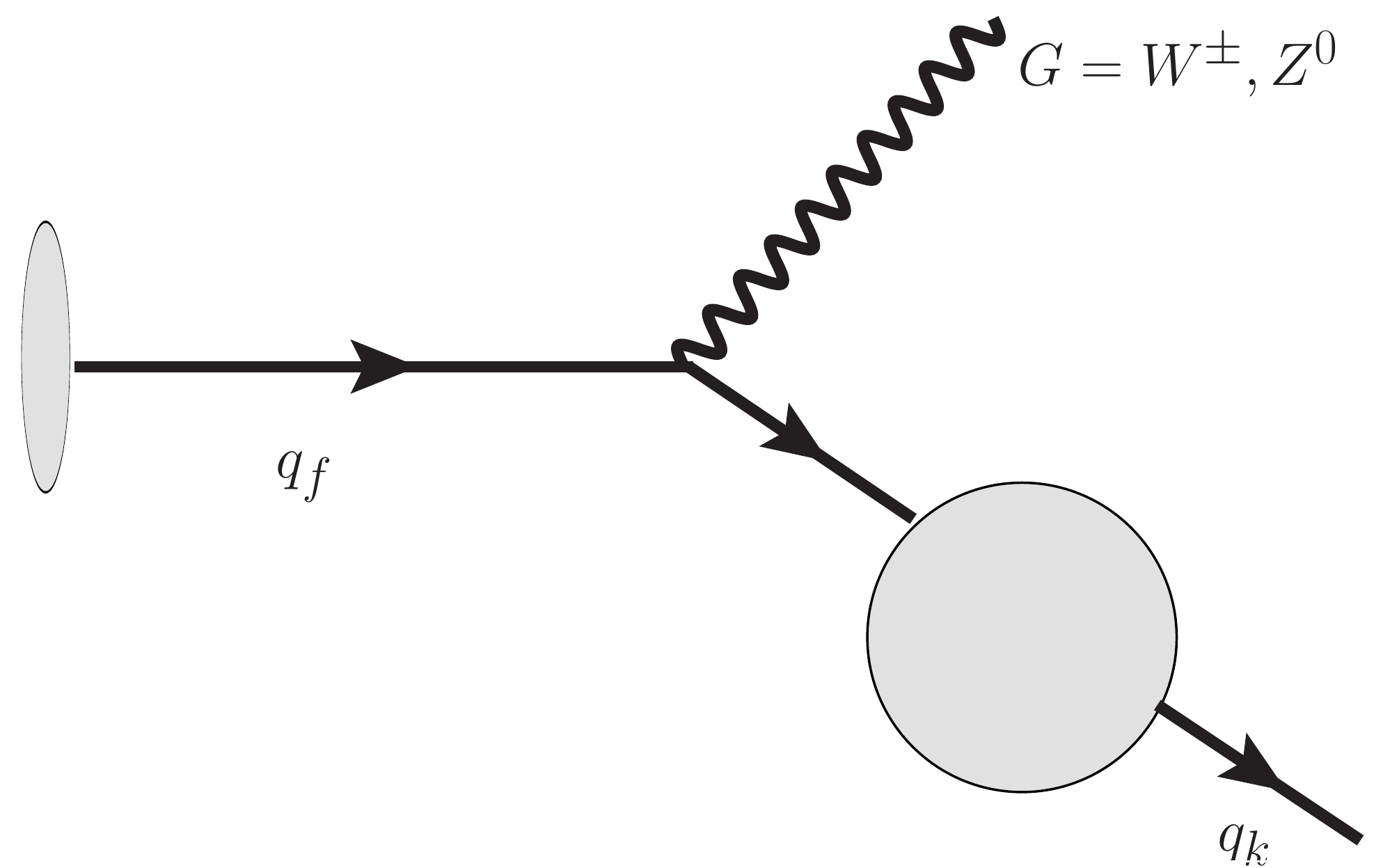}}
\label{fig:gb_frag}
}
\caption{Gauge boson radiation by a quark after (a) and before (b) 
the interaction with the target color field denoted by a shaded circle.}
\label{fig:gb_dip}
\end{figure}

Natually, the observables such as total cross sections are Lorentz invariant. A partonic scattering picture of a given 
process, however, depends on the reference frame [\refcite{k95}]. In the framework of conventional parton model in the 
center-of-mass frame the dilepton production emerges due the quark-antiquark annihilation into a virtual $\gamma/Z^0$ 
boson. In the target rest frame relevant for the dipole model formalism, however, the same process should be considered 
as a bremsstrahlung of $\gamma/Z^0$ and the corresponding diagrams are illustrated in Figs.~\ref{fig:gb_dip} (a) and (b), 
rather than $q\bar q$ annihilation [\refcite{kst99,hirr}]. Indeed, the gauge boson can be radiated off an energetic projectile quark
both before and after its scattering off the hadronic target such that these contributions interfere. Thus, at high energies
the incoming projectile quark effectively probes the dense gluonic field in the target nucleon and is particularly sensitive
to the nonlinear effects in multiple dipole-target scatterings. This is in analogy to the inclusive DIS reaction where a virtual 
photon in the Bjorken frame is considered as a probe for partonic structure of the hadron, while in the target rest frame 
it is instead viewed as an interaction of partonic components of the projectile photon.

It is well-known that soft fluctuations with a large sizes $\langle R^2 \rangle\sim 1/m_q^2$ ($m_q$ is a constituent light quark mass) 
corresponding to the asymmetric dipoles play a leading role in the diffractive DIS process, which is in variance with the inclusive 
DIS one [\refcite{rev1,k-povh}]. Indeed, even though the soft fluctuations are rather rare, they interact with a large cross section which
in practice compensates their tiny weights $\sim m_q^2/Q^2$. The hard fluctuations corresponding to symmetric small dipoles with
$\langle r^2 \rangle\sim 1/Q^2$ are much more abundant but their scattering is vanishing as $1/Q^2$. While the inclusive DIS is
a semi-hard/semi-soft process with the total cross section $1/Q^2$, the diffractive DIS is solely dominated by soft fluctuations 
$\sim 1/m_q^2Q^2$ leading to nearly $Q^2$-independence of the SD-to-inclusive ratio $\sigma_{\rm sd}/\sigma_{\rm inc}$ with 
the higher-twist behaviour of diffractive DIS.

The inclusive dilepton production process mediated by a virtual photon in the dipole framework in $pp$ and $pA$ collisions has 
been intensively studied in the literature so far (see e.g. Refs.~[\refcite{dynuc,rauf,gay,basso}]). In particular, it has been understood 
that the phenomenological dipole model predictions for DY observables are practically indistinguishable from those in the NLO collinear 
factorisation framework [\refcite{rauf}] and provide a good agreement with the recent LHC data [\refcite{basso}]. In the diffractive channel, 
the DY and electroweak gauge boson production has been studied within the dipole formalism in Ref.~[\refcite{WZ}].

The dipole formula for the inclusive DY cross section is similar to the DIS one [\refcite{hirr,kst99}]
\beq
\frac{d\sigma_{\rm inc}^{\rm DY}(qp\to\gamma^*X)}{d\alpha\,dM^2} =
\int d^2r\,\left|\Psi_{q\gamma^*}(\vec r,\alpha)\right|^2\,\sigma_{\bar qq}\left(\alpha r,x_2\right),
\label{60}
\eeq
in terms of $q\to q\gamma^*$ distribution function $\Psi_{q\gamma^*}(\vec r,\alpha)$ where $\alpha=p^+_{\gamma^*}/p^+_q$ 
is the fractional light-cone momentum of the virtual photon. The inclusive DIS and DY processes are related by means of QCD factorisation,
and a similarity between them is a source of the hadron PDF universality.

\begin{figure*}[!t]
\centering
 \includegraphics[width=12cm,clip]{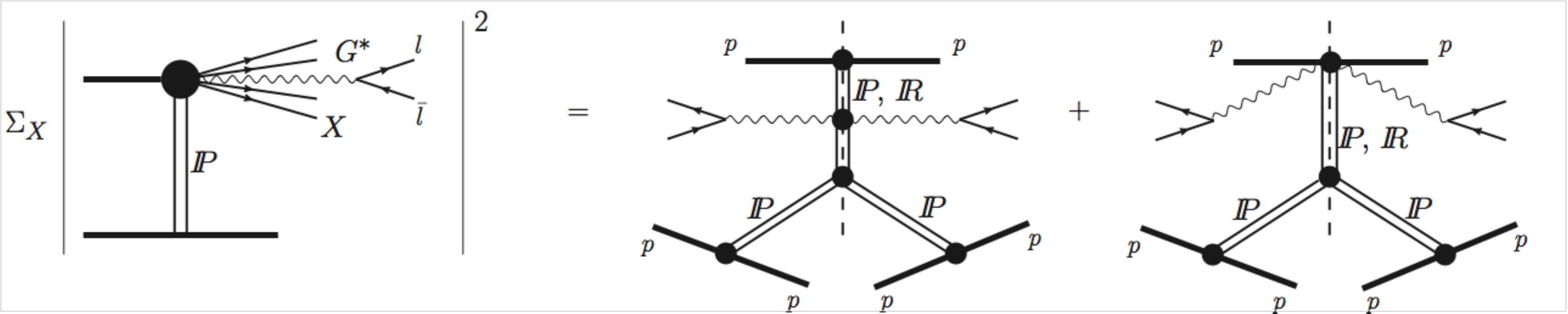}
\caption{Triple-Regge description of the process $pp\to Xp$, where the
diffractively produced state $X$ contains a gauge boson decaying to
a lepton pair.}
\label{fig:3-regge}      
\end{figure*}

The single diffractive (SD) DY process is characterized by a relatively small momentum transfer between the colliding protons, i.e.
both transverse and fractional momenta are small. One of the protons is then treated as a target, another -- as a projectile which
emits a virtual photon (or any other Abelian particle) and then hadronises into a hadronic system $X$ in forward region. Both $X$ 
and the radiated photon are separated by a large rapidity large from the target which remains intact, i.e.
\begin{eqnarray}
p_1+p_2\to X+(gap)+p_2 \,, \qquad X\equiv\gamma^*(l^+l^-)+Y \,.
\end{eqnarray}
At large Feynman $x_F\to1$ of the recoil target proton, the diffractive DY process is described by the triple Regge graphs 
illustrated in Fig.~\ref{fig:3-regge}. 

\begin{figure*}[!b]
\centering
 \includegraphics[width=12cm,clip]{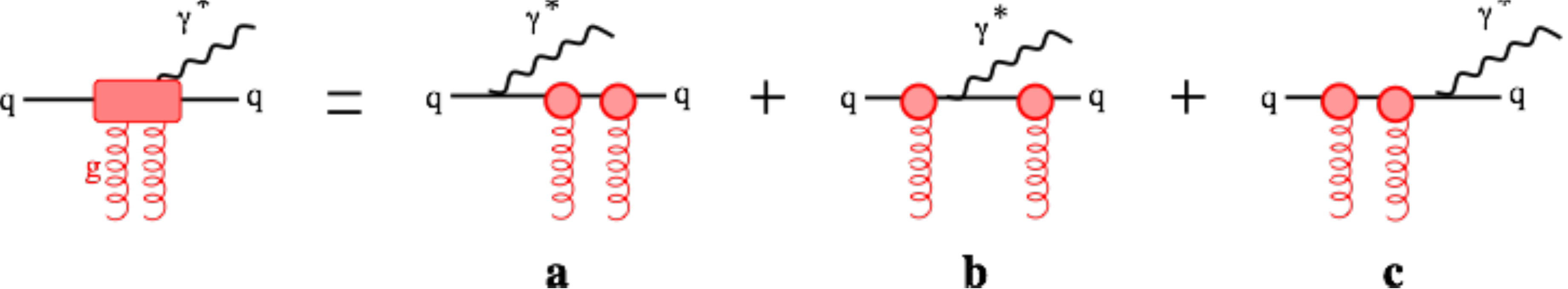}
\caption{Diffractive radiation of a virtual photon by a quark.}
\label{fig:dy-diff}       
\end{figure*}

In fact, diffractive DY process vanishes in the forward direction~[\refcite{kst99}] since the graphs (a), (b) and (c) 
shown in Fig.~\ref{fig:dy-diff} get canceled, i.e.
\beq
\left.\frac{d\sigma_{\rm inc}^{\rm DY}(qp\to\gamma^*qp)}  
{d\alpha\,dM^2\,d^2p_T}\right|_{p_T=0} = 0 \,.
\label{80}
\eeq
Indeed, only quark interacts with the target in its both Fock components, $|q\ra$ and $|q\gamma^*\ra$, so they interact with an equal
strengths and thus cancel in the forward diffractive amplitude in accordance to the Good-Walker picture. The same behavior holds for any 
diffractive Abelian radiation, in particular, for diffractive production of $\gamma^*$, $W^\pm$, $Z^0$ bosons as well as the Higgs boson.

The diffractive (Ingelman-Schlein) factorisation in the diffractive Abelian radiation with a large rapidity gap is broken. Indeed, large- and small-size 
fluctuations in the projectile cannot be consistently separated and contribute to the diffractive Abelian radiation on the same footing. This is the source
of the leading-twist behavior of the diffractive DY cross section whereas the diffractive DIS is determined by soft fluctuations and thus 
emerges as a higher-twist process [\refcite{KPST-06,dy-r}]. 
\begin{figure*}[!t]
\vspace*{-1.5cm}
\begin{minipage}{0.495\textwidth}
 \centerline{\includegraphics[width=1.0\textwidth]{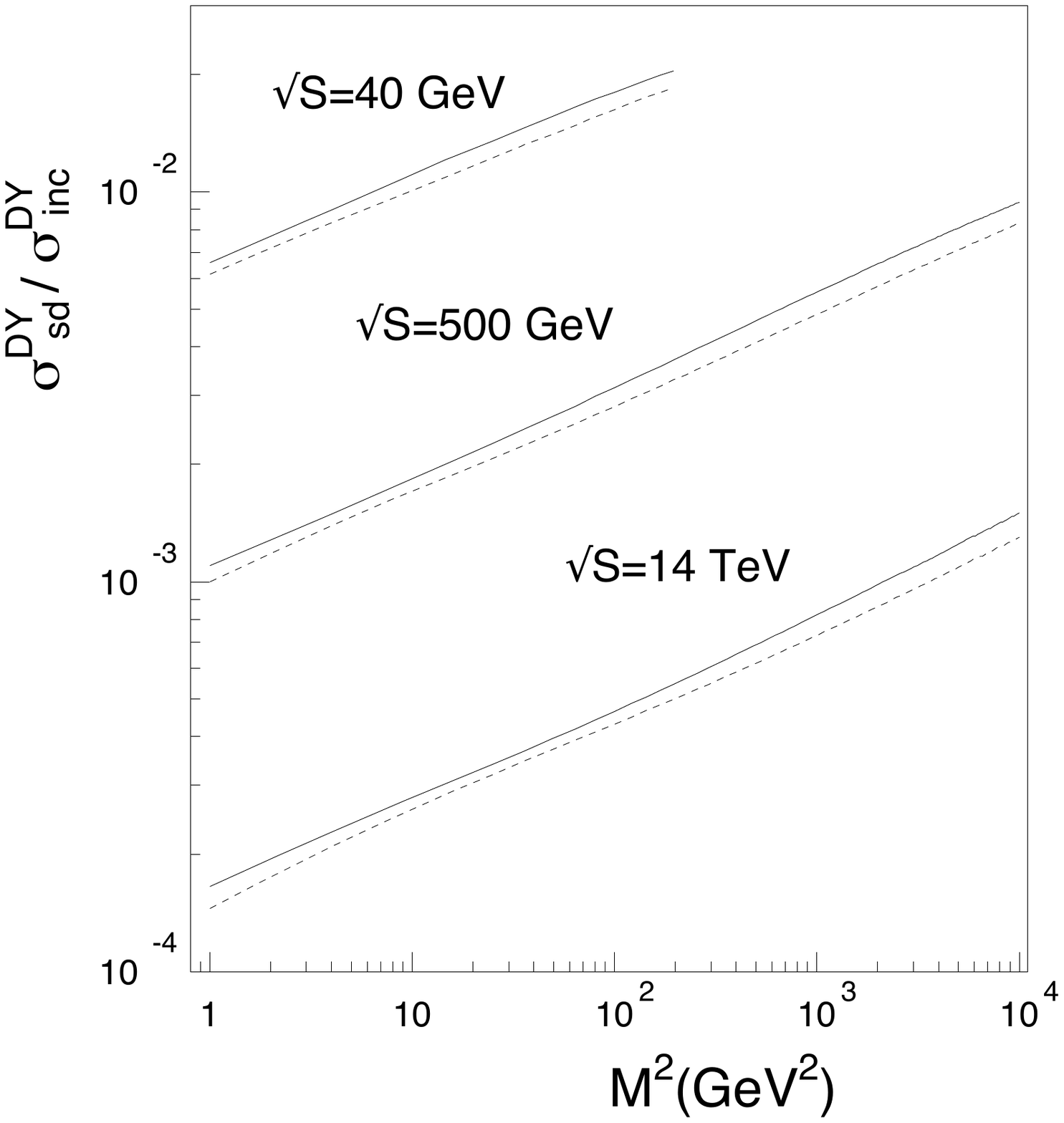}}
\end{minipage}
\begin{minipage}{0.495\textwidth}
 \centerline{\includegraphics[width=1.0\textwidth]{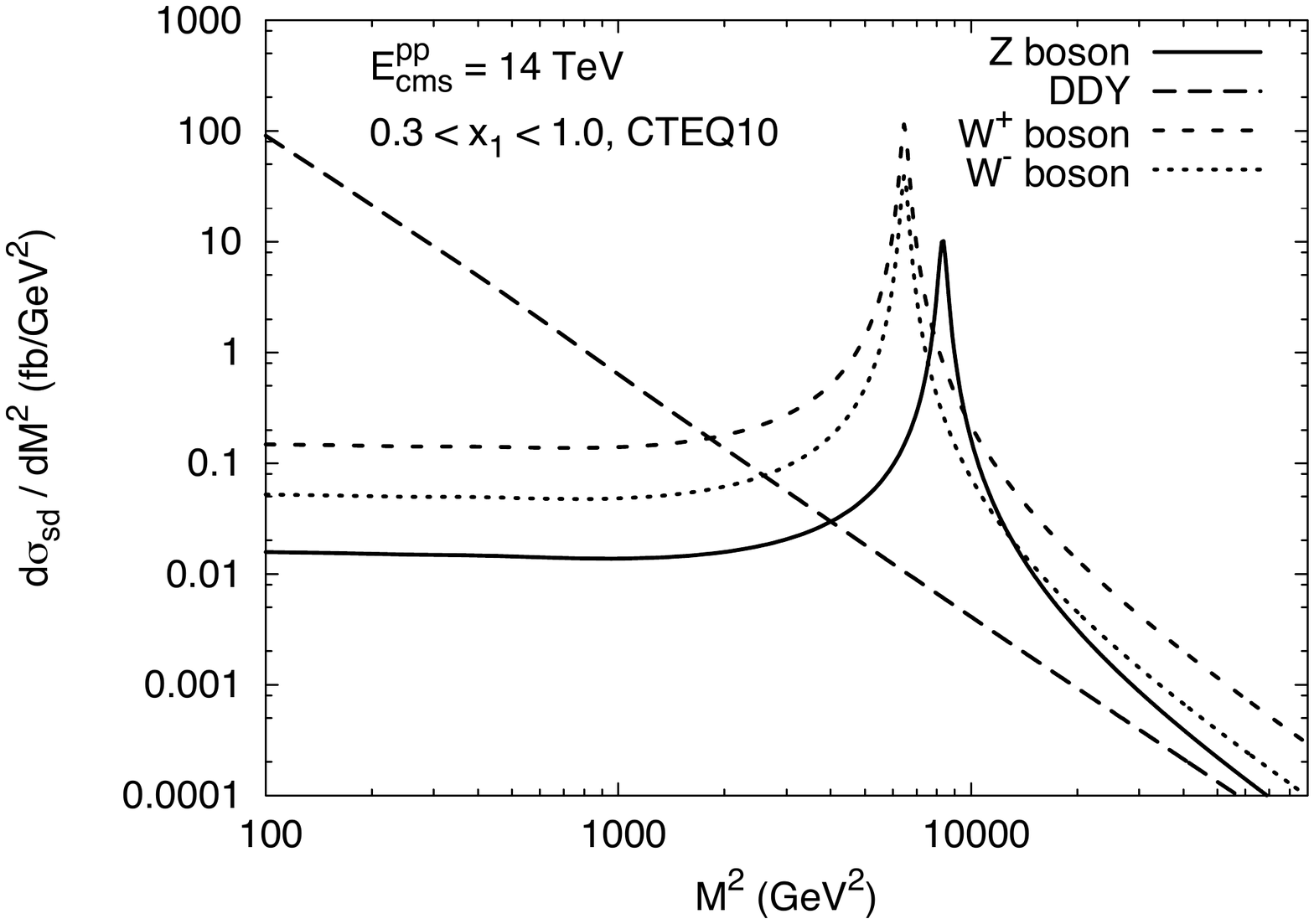}}
\end{minipage}
\vspace*{-1.5cm}  
   \caption{ \small 
The SD-to-inclusive ratio of the DY cross sections vs the dilepton mass squared for c.m. energies $\sqrt{s}=0.04,0.5$ 
and $14$ TeV (left panel) and the diffractive DY and gauge $W^\pm$, $Z^0$ boson production cross sections as functions 
of dilepton invariant mass squared $M^2$ at $\sqrt{s}=14$ TeV (right panel). }
\label{fig:dy-results}
\end{figure*}

Due to the internal transverse motion of the projectile valence quarks inside the incoming proton, which corresponds 
to finite large transverse separations between them, the forward photon radiation does not vanish [\refcite{KPST-06,dy-r}].
These large distances are controlled by a nonperturbative (hadron) scale $\vec R$, such that the diffractive amplitude 
has the Good-Walker structure (\ref{40}), such that the ratio of the cross sections reads
\beq
\frac{\sigma_{\rm sd}^{\rm DY}}{\sigma_{\rm incl}^{\rm DY}}\propto
\left[\sigma_{\bar qq}(\vec R+\vec r,x_2)-\sigma_{\bar qq}(\vec R,x_2)\right]^2\propto
\frac{\exp(-2R^2/R_0^2(x_2))}{R_0^2(x_2)} \,.
\label{100}
\eeq
\begin{figure*}[!b]
 \centerline{\includegraphics[width=0.6\textwidth]{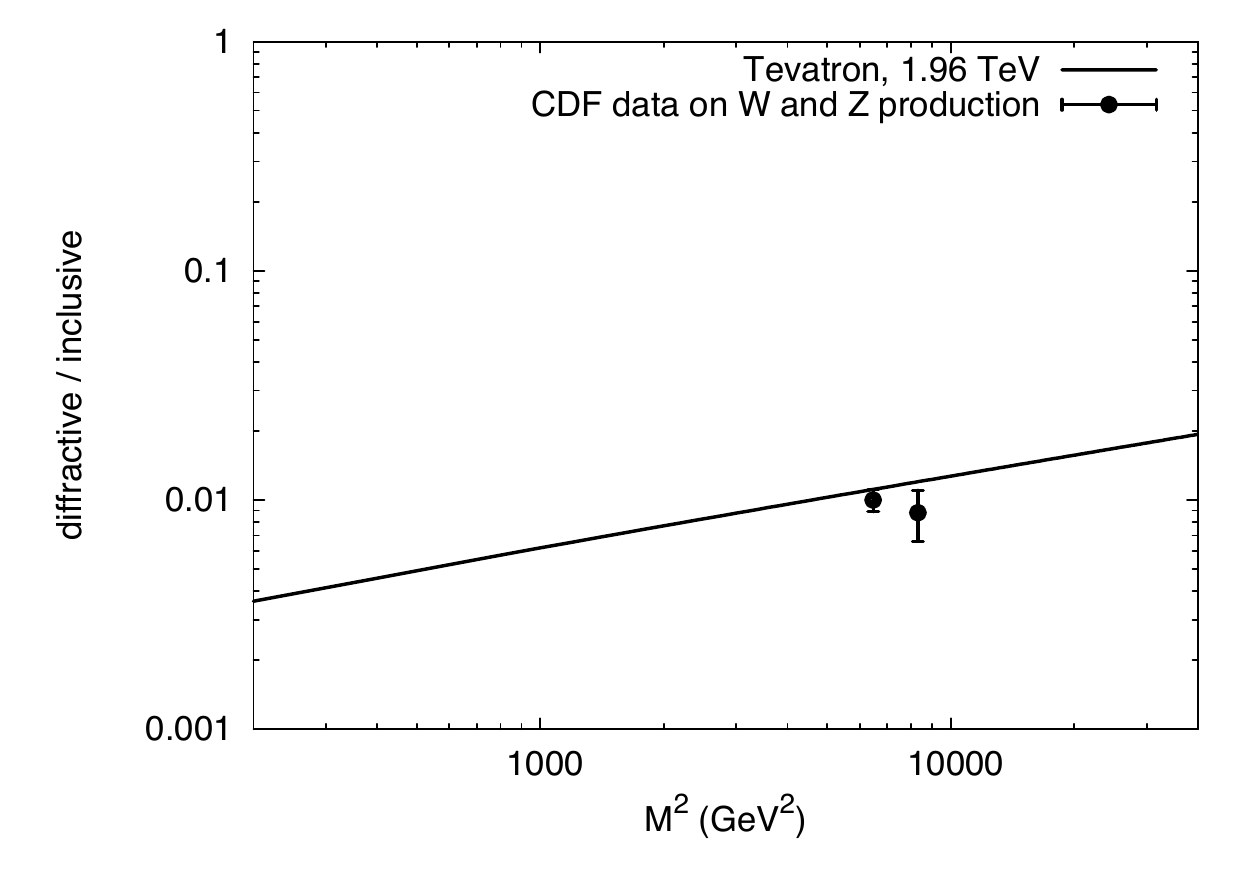}}
   \caption{
 The SD-to-inclusive ratio vs dilepton invariant mass squared in comparison with the CDF data [\citen{Aaltonen:2010qe}].
 }
 \label{fig:diff-data}
\end{figure*}

Thus, the soft part of the diffractive DY interaction is not enhanced such that the process is semi-hard/semi-soft 
similarly to the inclusive DIS one. The emergent linear dependence on the hard scale $r\sim 1/M \ll 
R_0(x_2)$ indicates that even at a hard scale the diffractive Abelian radiation is very sensitive to the hadron scale,
thus, diffractive factorisation does not hold and should not be imposed in practice [\refcite{Landshoff-DDY}]. The observation
that factorisation in diffractive DY process fails due to the presence of spectator partons in the Pomeron has been 
first made in Refs.~[\refcite{Collins93,Collins97}]. Recently, in Refs.~[\refcite{KPST-06,dy-r,WZ}] it was explicitly shown 
that factorisation in diffractive Abelian radiation is therefore even more broken due to presence of spectator 
partons in the colliding hadrons. Besides, the saturated shape of the dipole cross section (\ref{Sig-dip}), therefore, 
leads to several unusual features of diffractive DY cross section. Namely, the fractional diffractive DY cross section 
is steeply falling with energy, but rises with the scale as is demonstrated in Fig.~\ref{fig:dy-results}.

In general, diffractive radiation of any Abelian particle is given by the same graphs as the diffractive DY process, with an
appropriate use of couplings and spin structure [\refcite{WZ}]. In Fig.~\ref{fig:dy-results} (right panel) we show the single 
diffractive cross sections for $Z^0,\,\gamma^*$ and $W^{\pm}$ bosons production as functions of the dilepton mass 
squared, $d\sigma_{\rm sd}/dM^2$. The SD-to-inclusive ratios of the DY cross sections for diffractive $Z^0$ and $W^\pm$ 
production are shown in Fig.~\ref{fig:diff-data} in comparison with the CDF data [\refcite{Aaltonen:2010qe}].

\section{Gap survival effect at the amplitude level}
\label{Sec:GS}

A suppression factor in diffractive cross sections which parameterises the unitarity corrections is known as the survival probability. 
The latter significantly reduces the diffractive observables in hadronic collisions and there is no process-independent way to compute 
it consistently beyond the Regge theory. The soft survival probability emerges due to the long-range interactions of (soft) spectator partons
in the projectile hadron wave function which are not present in the case of diffractive DIS. Hence, the transverse motion of spectators is
the basic source of diffractive factorisation breaking in hadronic collisions compared to diffraction in $ep$ collisions.
\begin{figure}[!h]
\centerline{\epsfig{file=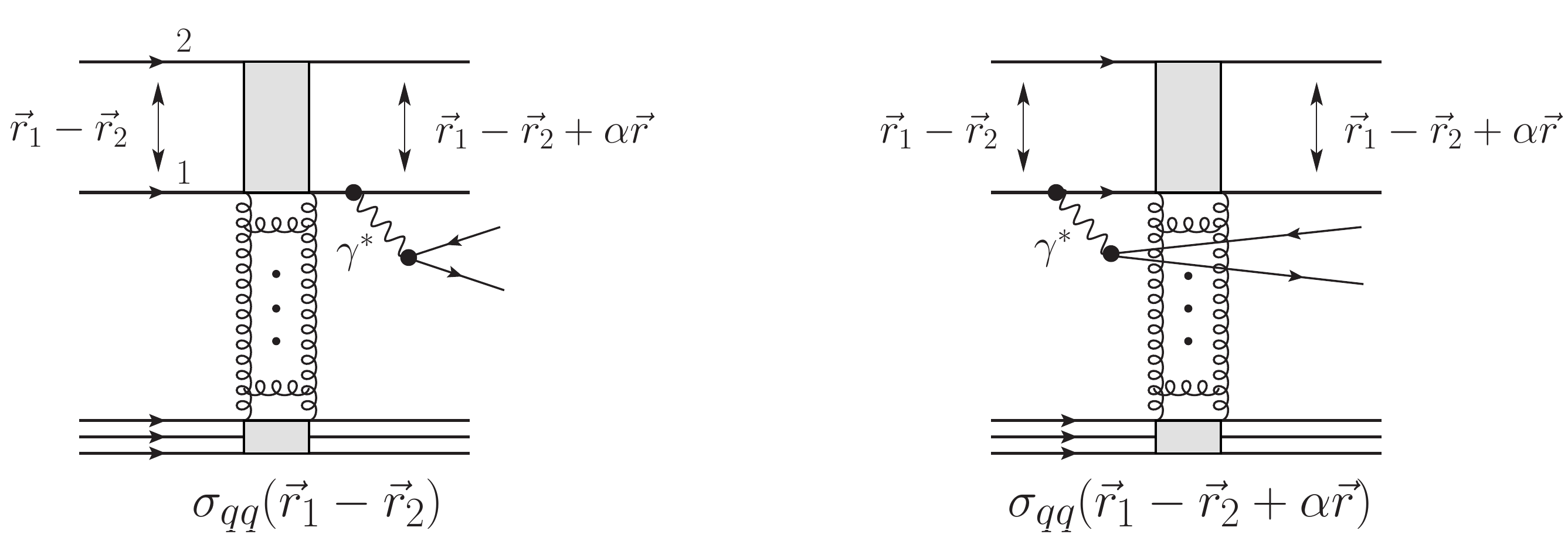,width=12cm}} 
 \caption{Diffractive photon radiation in the dipole-target scattering as an elementary ingredient of diffractive DY process 
 in hadronic collisions.}
\label{fig:gam}
\end{figure}

The forward diffractive DY process in hadronic collisions is dominated by a large-size dipole elastic scattering off a given potential 
as illustrated in Fig.~\ref{fig:gam}. Let $\vec{r}_{1,2}$ are the positions of projectile quarks in the transverse plane of the hadronic wave function 
such that the dipole size is $|\vec r_1 - \vec r_2| \sim R_{\rm had}$. As was mentioned above, a quark emitting a virtual photon shifts its position 
by a value determined by the hard scale $\vec \delta = \alpha \vec r \sim 1/M \ll R_{\rm had}$ such that the dipole size is changed. As was discussed above, the 
interference between the two graphs in Fig.~\ref{fig:gam} leads to a non-vanishing contribution to the diffractive DY cross section. The dipole partial 
amplitude can then be written in the eikonal form as
\begin{equation}
 \mathrm{Im}\,f_{\rm el}(\vec{b},\vec{r}_1-\vec{r}_2)= 1 - \exp\Big(i{\cal V}(\vec r_1) - i{\cal V}(\vec r_2)\Big) \,, \qquad 
 {\cal V}(b)=-\int\limits_{-\infty}^\infty dz\,V(\vec b,z) \,,
\label{model}
\end{equation}
where the scattering potential is denoted as $V(\vec b,z)$. At high energies, this amplitude is nearly imaginary
such that the diffractive DY amplitude off a dipole is provided by
\begin{equation*}
\mathrm{Im}\,f_{\rm el}(\vec{b}, \vec{r}_1-\vec{r}_2+\alpha\vec r) - \mathrm{Im}\,f_{\rm el}(\vec{b}, \vec{r}_1-\vec{r}_2)\simeq
\exp\Big(i{\cal V}(\vec r_1)-i{\cal V}(\vec r_2)\Big)\, \exp\Big(i\alpha\,\vec r\cdot\vec\nabla{\cal V}(\vec r_1)\Big)\,,
\end{equation*}
where the first exponential factor represents the gap survival amplitude which accounts for all absorptive corrections, provided that the universal 
dipole cross section is fitted to the data and accounts for both hard and soft interactions. The survival amplitude indeed vanishes in the black disk 
asymptotics as required. While conventionally the gap survival factor is incorporated directly into the diffractive cross sections and is thus treated 
probabilistically, in the color dipole framework the corresponding effects are accounted automatically and treated more naturally quantum-mechanically.

\section{Diffractive heavy flavor production}
\label{Sec:hf}

Understanding of both inclusive and diffractive hadroproduction of heavy quarks at large Feynman $x_F\to 1$ is a longstanding 
controversial problem. Indeed, QCD factorisation predicts vanishing $Q\bar Q$ production cross sections at large $x_F$ due to a steeply
decreasing gluon density in the forward kinematics which contradicts to the end-point behavior predicted by the Regge
asymptotics (see e.g. Ref.~[\refcite{rev2}] and references therein). A similar contradiction arises for the DY reaction at large
$x_F$ which was seen from the data [\refcite{Wijesooriya:2005ir}]. Both examples apparently indicate that the conventional
QCD factorisation does not hold, at least, at large Feynman $x_F$ [\refcite{Kopeliovich:2005ym}].
\begin{figure*}[!h]
\centering
 \includegraphics[width=6cm,clip]{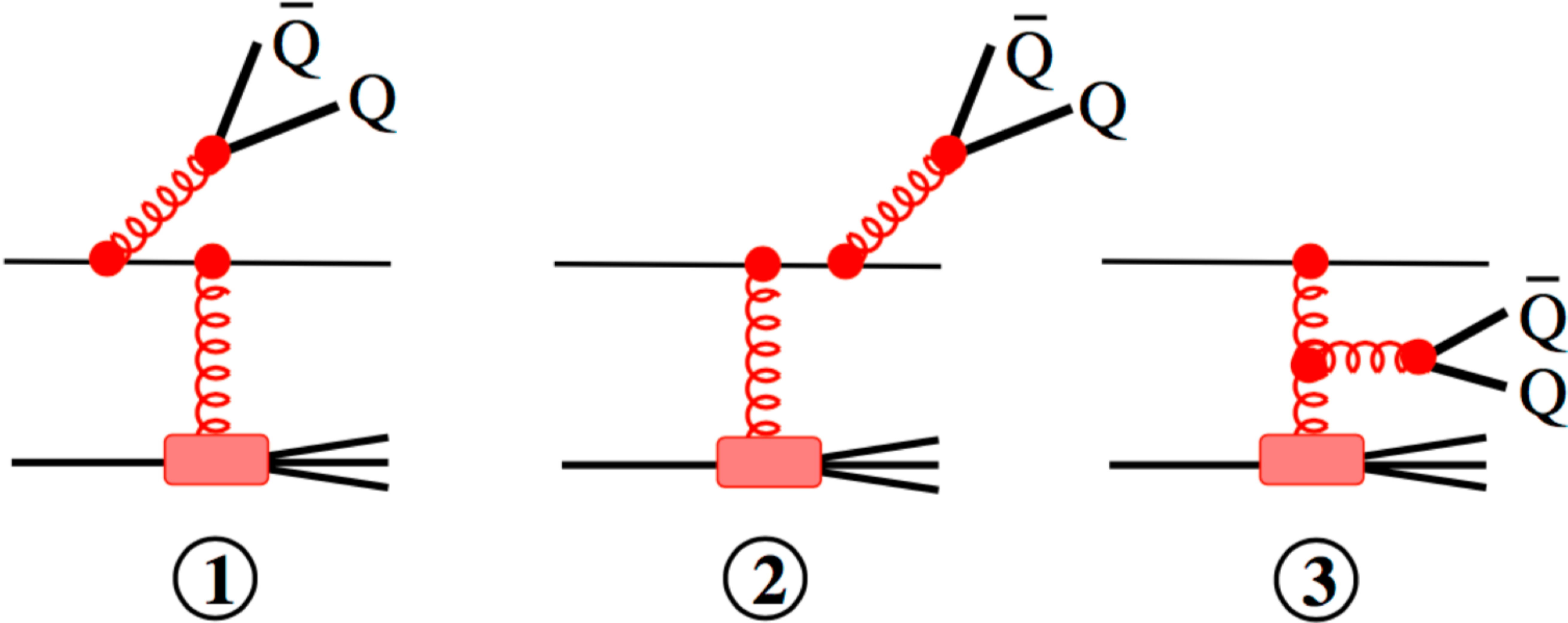}
 \hspace*{5mm}
 \includegraphics[width=3.5cm,clip]{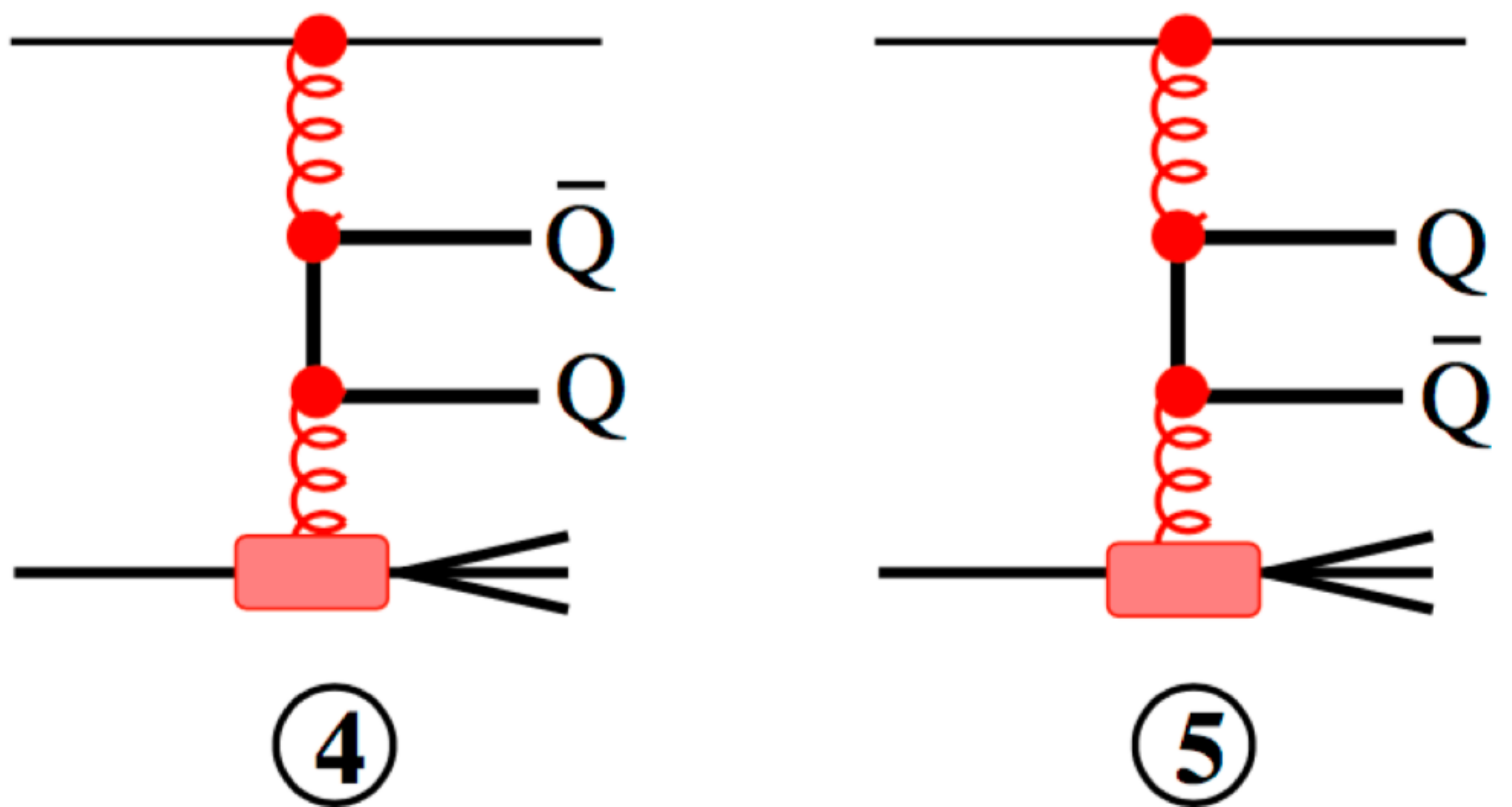}
\caption{Typical contributions to inclusive production of a heavy quark pair in a quark-proton collision.}
\label{fig:graphs}      
\end{figure*}

A detailed analysis of various contributions into the diffractive $Q\bar Q$ production from both diffractive gluon and quark excitations
has been performed in Ref.~[\refcite{hf-diff}]. For example, in the case of diffractive quark excitation $q+g\to (Q\bar Q)+q$ the dynamics 
of inclusive heavy flavor production is characretized by five distinct topologies which can be classified as: (i) bremsstrahlung (like in DY), 
and (ii) production mechanisms as illustrated by the Feynman graphs in Fig.~\ref{fig:graphs}, such that the total amplitude
 \beq
A^{\rm diff}_{Q\bar Q} = A_{\rm BR} + A_{\rm PR}\,.
 \eeq
Each of these two contributions is gauge invariant and can be described in terms of three-body dipole cross sections, $\sigma_{g\bar qq}$ 
and $\sigma_{g\bar QQ}$, respectively, which strongly motivates such a separation. Similar graphs and classification hold for the diffractive 
gluon excitation $g+g\to (Q\bar Q)+g$ as well. The amplitudes for each of the two mechanisms are expressed via the amplitudes $A_i$ 
corresponding to the graph numbering in Fig.~\ref{fig:graphs}. As was elaborated in Ref.~[\refcite{hf-diff}] such a grouping can be performed 
for both transversely and longitudinally polarised indermediate gluons. The bremsstrahlung and production components have the following form,
\beqn
A_{\rm BR}&=&A_1+A_2+\frac{Q^2}{M^2+Q^2}\,A_3 \,;
\label{120}
\\
A_{\rm PR}&=&\frac{M^2}{M^2+Q^2}\,A_3+A_4+A_5 \,,
\label{140}
\eeqn
where $Q^2=(p_i-p_j)^2$ in terms of the initial $p_i$ and final $p_j$ projectile quark momenta, 
and $M$ is the invariant mass of the $Q\bar Q$ pair.
\begin{figure*}[!t]
\centering
 \includegraphics[width=4cm,clip]{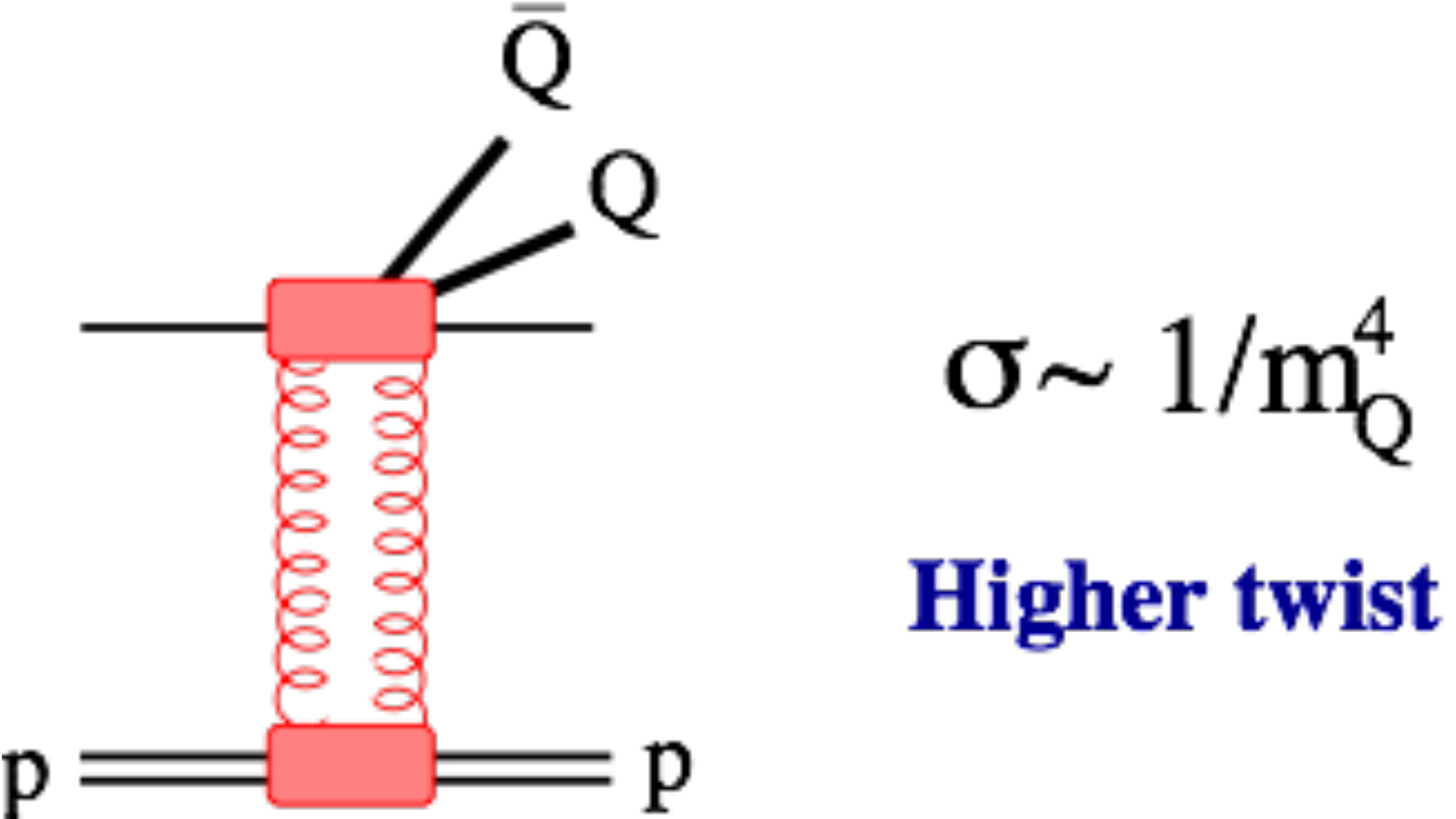}
 \hspace*{1cm}
 \includegraphics[width=7cm,clip]{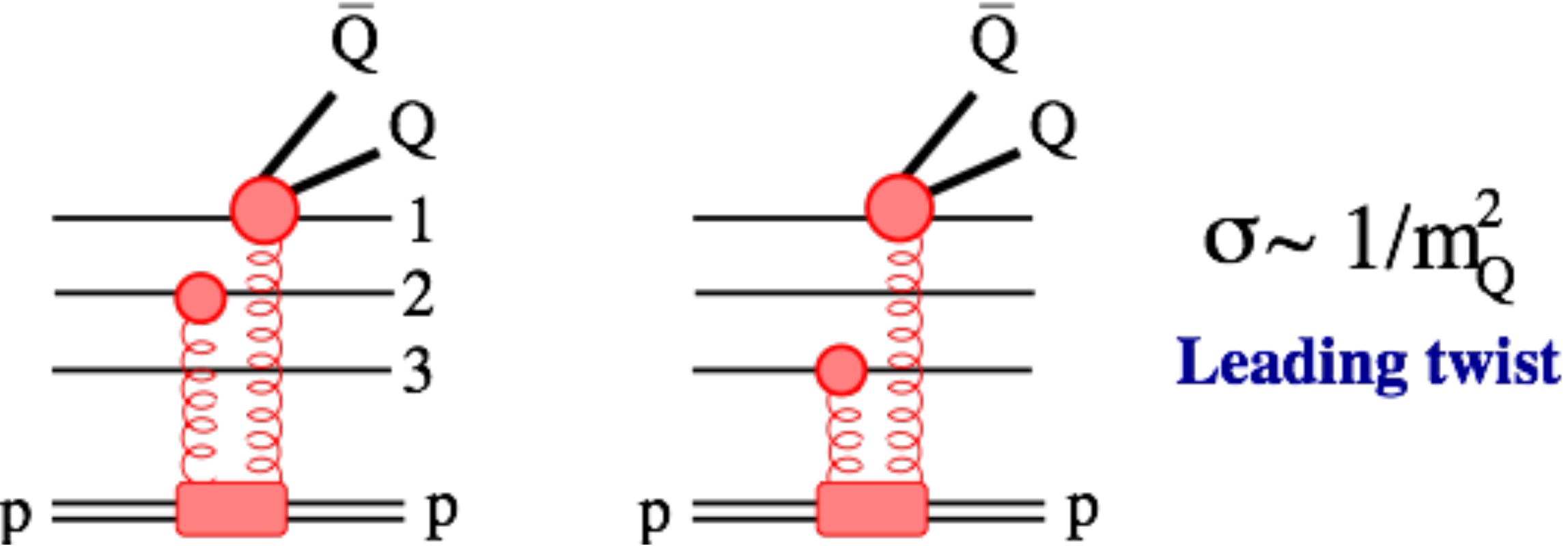}
\caption{Diffractive production of a heavy quark pair in a quark-proton collision (left panel) and 
in a proton-proton collision (right panel).}
\label{fig:twists}      
\end{figure*}

For diffractive production one has to provide a colorless two-gluon exchange. In analogy to the leading-twist DIS diffraction at large photon virtualities 
$\gamma^*\to Q\bar Qg$, the BR and PR contributions are dependent on two characteristic length scales: the small separation between the $\bar Q$ and $Q$, 
$s\sim 1/m_Q$, and a typically large separation between $q$ and $Q \bar Q$, $\rho\sim 1/m_q$. In analogy to diffractive DY, the diffractive 
excitation of a quark thus turns out to be a higher twist effect as is depicted in Fig.~\ref{fig:twists} (left). The leading twist contributions to 
diffractive $Q\bar Q$ production come from both sources: when both exchanged gluons couple to the valence quark which gives rise to the 
$Q\bar Q$ pair, and when one of the gluons is coupled to another spectator quark not participating in the hard scattering as is shown in 
Fig.~\ref{fig:twists}, right [\refcite{hf-diff}] (for more details, see Ref.~[\refcite{rev2}]). So the interaction with spectators again plays an important role
as one of the source for the diffractive factorisation breaking.
\begin{figure*}[!b]
\begin{minipage}{0.495\textwidth}
 \centerline{\includegraphics[width=0.7\textwidth]{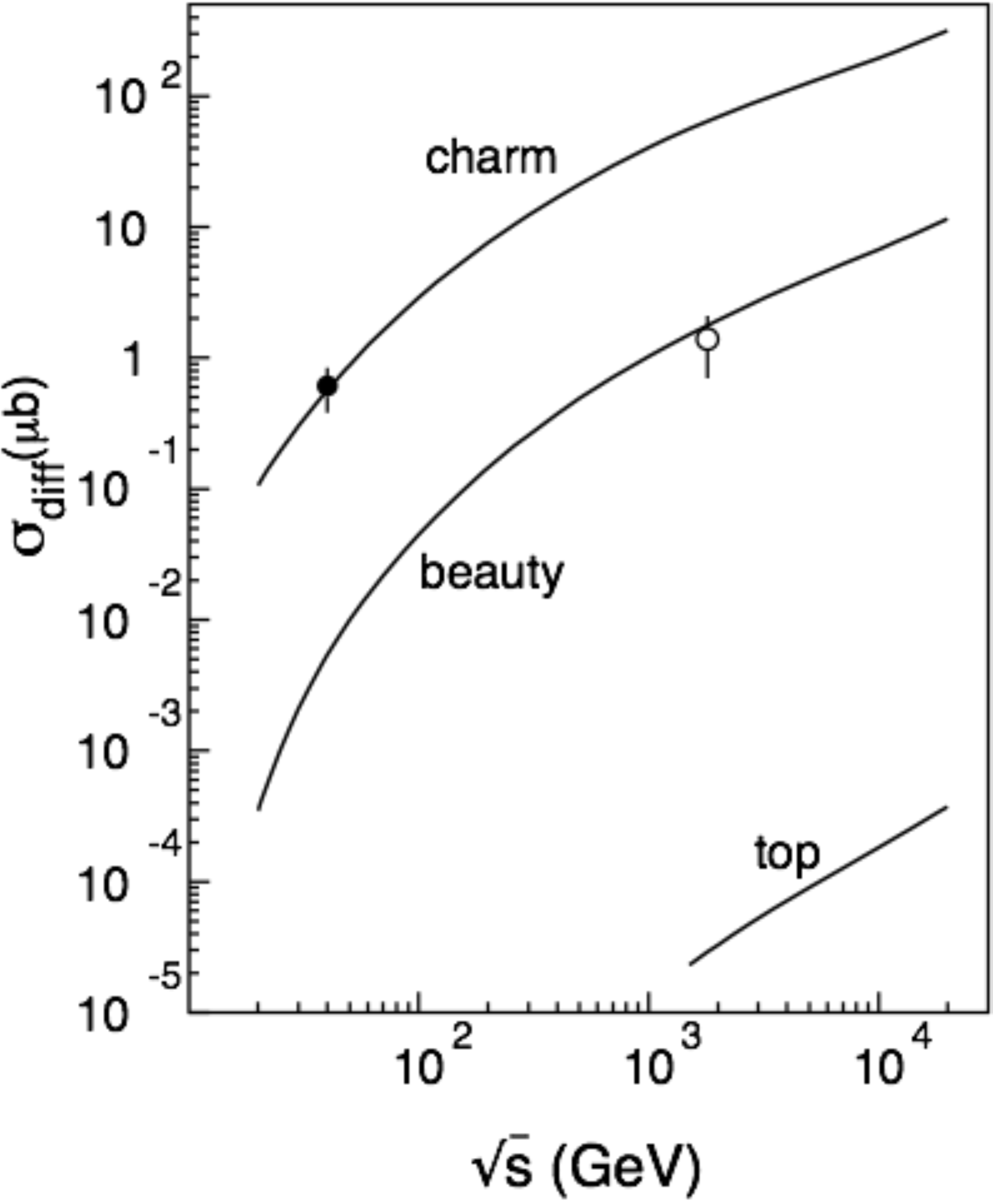}}
\end{minipage}
\begin{minipage}{0.495\textwidth}
 \centerline{\includegraphics[width=0.72\textwidth]{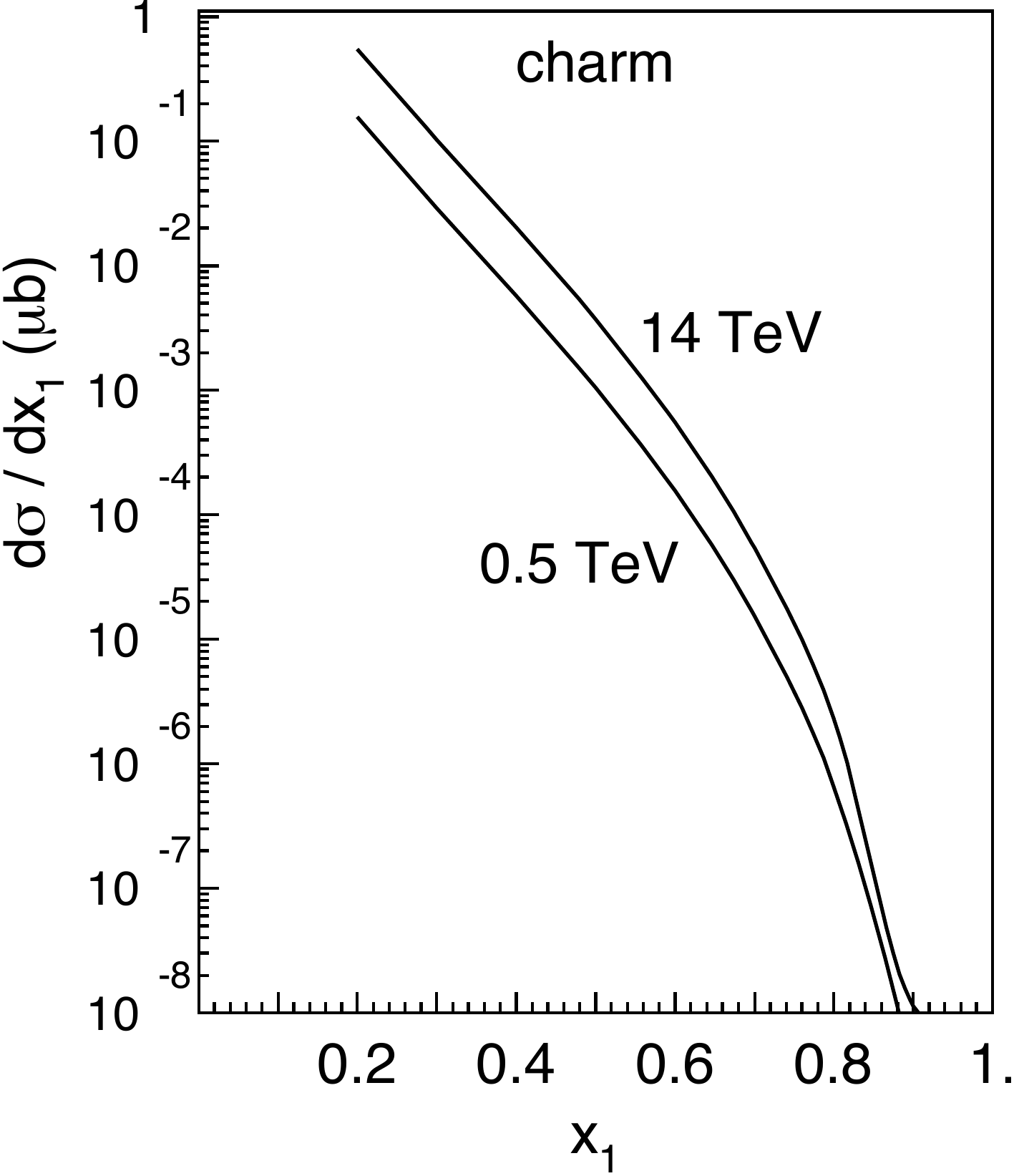}}
\end{minipage}
   \caption{The total cross sections for diffractive heavy flavor production w.r.t. 
the experimental data from E690 [\citen{Wang:2000dq}] and CDF [\citen{Affolder:1999hm}] 
experiments for charm and beauty as functions of energy and differential cross section $d\sigma/dx_1$ for 
diffractive charm quark production at $\sqrt{s}=0.5$ and 14 TeV.}
\label{fig:diff-QQ}
\end{figure*}

The $Q\bar Q$ production amplitudes in diffractive quark scattering off a proton target are related
to the effective dipole cross sections $\Sigma_{1,2}$ for colorless $g\bar qq$ and $g\bar QQ$ systems as
\begin{eqnarray*}
&& A_{\rm BR}\propto \Phi_{\rm BR}(\vec \rho,\vec s)\Sigma_1(\vec \rho,\vec s)\sim \langle s^2 \rangle\sim \frac{1}{m_Q^2}\,, \\
&& A_{\rm PR}\propto \Phi_{\rm PR}(\vec \rho,\vec s)\Sigma_2(\vec \rho,\vec s)\sim \vec s \cdot \vec \rho \sim \frac{1}{m_q m_Q} \,,
\end{eqnarray*}
where $\Phi_{\rm BR/PR}$ are complicated distribution amplitudes for the $q+g\to (Q\bar Q)+q$ subprocess. 
The bremsstrahlung contribution is of a higher twist effect and is therefore suppressed while for diffractive Abelian radiation
it is equal to zero. In opposite, the production contribution is of the leading twist and is thus much larger than the bremsstrahlung term 
in analogy to the diffractive DY reaction. This is again due to the presence of spectators at large distances from the $Q\bar Q$ pair despite
of non-Abelian nature of the process which is a rather non-trivial fact. The non-Abelian interactions, however, introduce extra important 
leading-twist terms into the ``production'' mechanism, which are independent of the structure of the hadronic wave function, in addition 
to those from the spectators' interactions. 

The leading-twist behavior $1/m_Q^2$ of the diffractive cross section is confirmed by E690 [\refcite{Wang:2000dq}] and CDF 
[\refcite{Affolder:1999hm}] data as demonstrated in Fig.~\ref{fig:diff-QQ} (left panel), where the corresponding cross sections 
for charm, beauty and top quarks, $p+p\to Q\bar QX+p$, are shown as functions of c.m.s. $pp$ energy. Besides,
on the right panel we show differential cross section in $x_1$-variable, $d\sigma/dx_1$, for diffractive charm quark production 
at two different energies $\sqrt{s}=0.5$ and 14 TeV.
\begin{figure*}[!h]
\centering
 \includegraphics[width=11cm,clip]{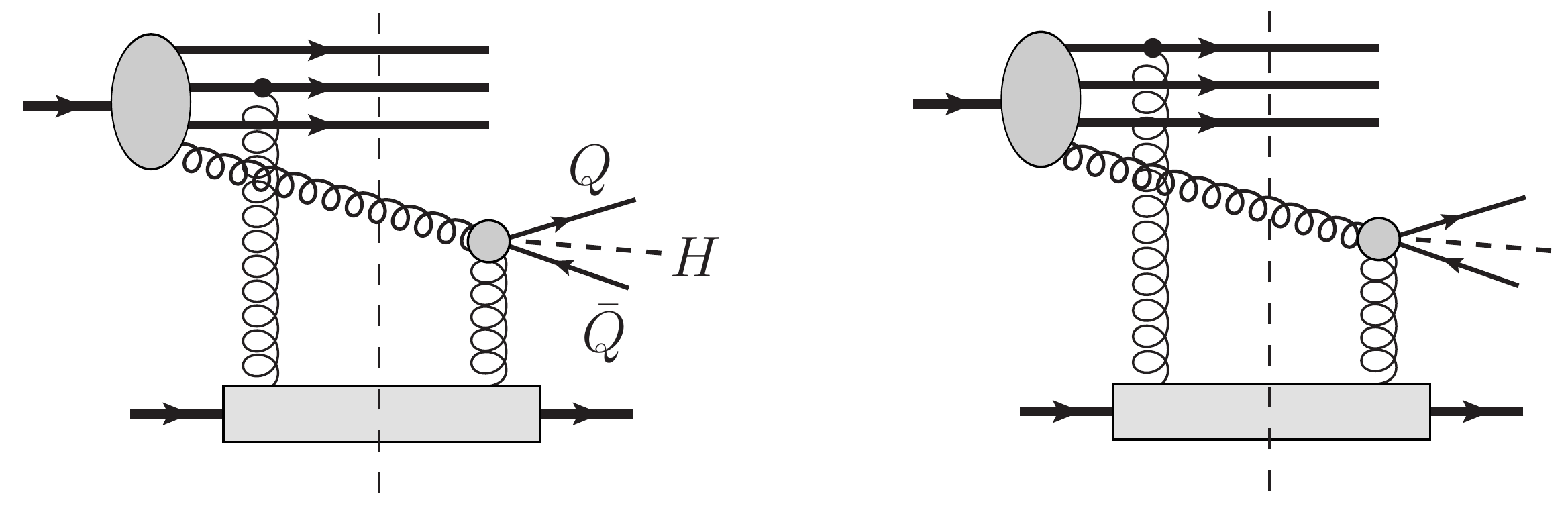}
\caption{Typical Feynman graphs for the diffractive Higgsstrahlung process off 
a heavy quark which involve interations of spectator partons.}
\label{fig:higgs-graphs}      
\end{figure*}

\section{Diffractive Higgs production}
\label{Sec:higgs}

\subsection{Higgsstrahlung}

Consider single diffractive Higgs boson production in hadron-hadron collisions. The Higgs boson decouples from light quarks, in particular, 
due to a smallness of the corresponding Yukawa coupling so the Higgsstrahlung by light hadrons is vanishingly small. Although a light projectile 
quark does not radiate the Higgs boson directly, it can do it via production of heavy flavors. Similarly to the diffractive $Q\bar Q$ production 
considered above, the diffractive Higgsstrahlung process off a heavy quark is dominated by the diagrams involving interactions of spectators 
at large transverse separations as illustrated in Fig.~\ref{fig:higgs-graphs}. Therefore, the Higgsstrahlung mechanism is closely related to 
the non-Abelian mechanism for diffractive heavy quark production discussed in the previous section. In a sense, it is also similar to diffractive 
DY, $Z^0$ and $W^\pm$ production since in all these cases the radiated particle does not participate in the interaction with the target 
although $gg \to Q\bar Q+H$ subprocess is rather involved and more complicated Fock states containing heavy flavors need to be resolved
by the exchanged gluons. 
\begin{figure*}[!h]
\begin{minipage}{0.495\textwidth}
 \centerline{\includegraphics[width=1.0\textwidth]{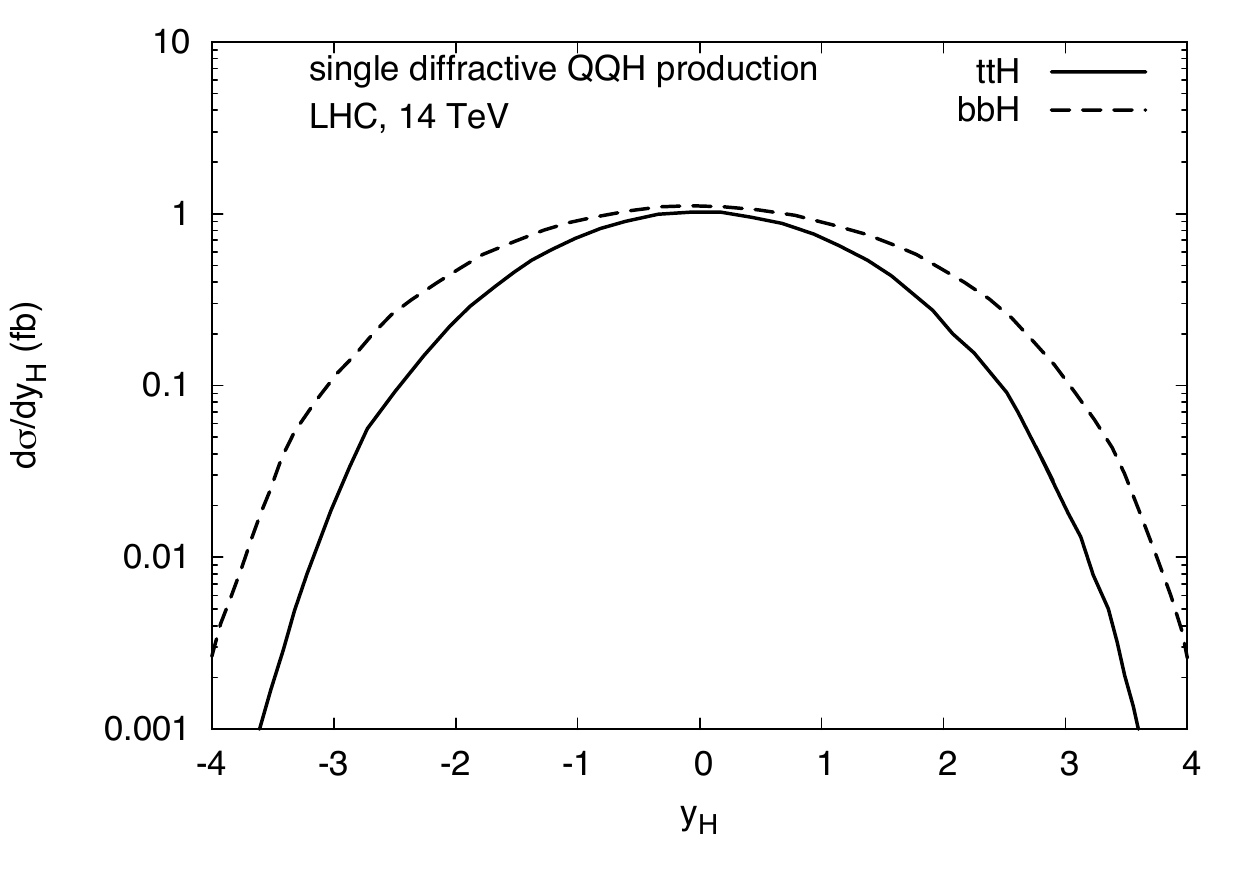}}
\end{minipage}
\begin{minipage}{0.495\textwidth}
 \centerline{\includegraphics[width=1.0\textwidth]{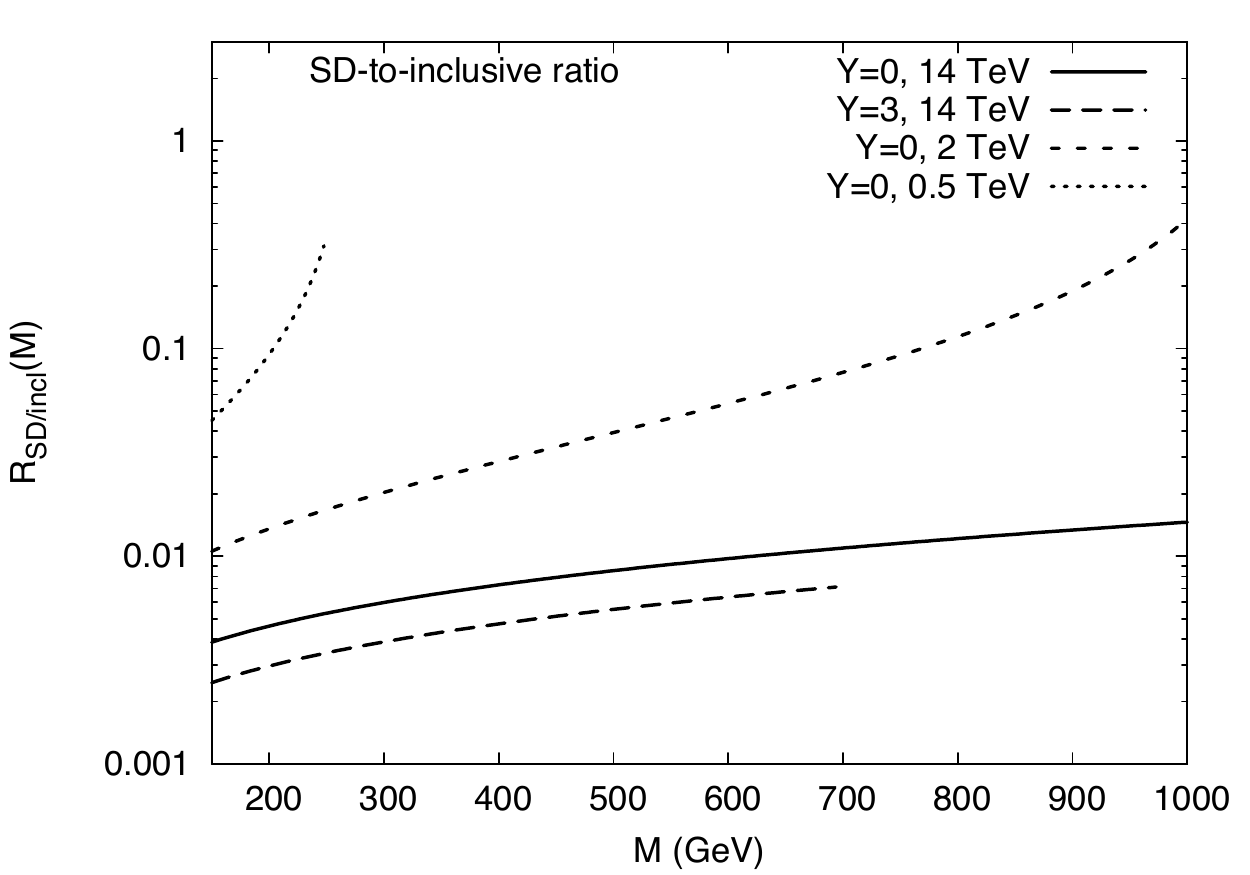}}
\end{minipage}
   \caption{The differential cross section of single diffractive Higgs boson 
production in association with a heavy quark ($b\bar b$ and $t\bar t$) pair 
vs Higgs boson rapidity (left panel) and the SD-to-inclusive ratio for the Higgsstrahlung
process as a function of the $Q\bar QH$ invariant mass (right panel) (see more 
details in Ref.~[\citen{higgs-r}]).}
\label{fig:higgs}
\end{figure*}

The rapidity-dependent cross section of diffractive Higgs boson production off $t\bar t$ and $b\bar b$ at the LHC energy 
$\sqrt{s}=14$ TeV is plotted in Fig.~\ref{fig:higgs} (left panel). At Higgs mid-rapidities, the top and bottom 
contributions are comparable to each other, whereas top quark provides a wider rapidity distribution and dominates at large
Higgs boson transverse momentum [\refcite{higgs-r}]. The total cross section is rather small and below $1$ fb. In Fig.~\ref{fig:higgs} 
(right panel) we present the SD-to-inclusive ratio of the corresponding Higgsstrahlung cross sections for different c.m. energies 
$\sqrt{s}=0.5,7,14$ TeV and for two values of the Higgs boson rapidities $Y=0$ and $3$ as functions of $\bar QQH$ invariant mass. 
This ratio is in overall agreement with the corresponding data for diffractive beauty production [\refcite{Affolder:1999hm}]. 

As expected from above discussion, the diffractive factorisation in diffractive Higgsstrahlung is broken by transverse motion 
of spectator valence quarks in the projectile hadron leading to a growth of the SD-to-inclusive ratio with the hard scale, $M$, and 
its descrease with $\sqrt{s}$. Such a behavior is opposite to the one predicted by diffractive factorisation and is in full analogy 
with the diffractive Abelian radiation.

\subsection{Direct heavy flavour fusion}

\begin{figure*}[!h]
\centering
 \includegraphics[width=6.5cm,clip]{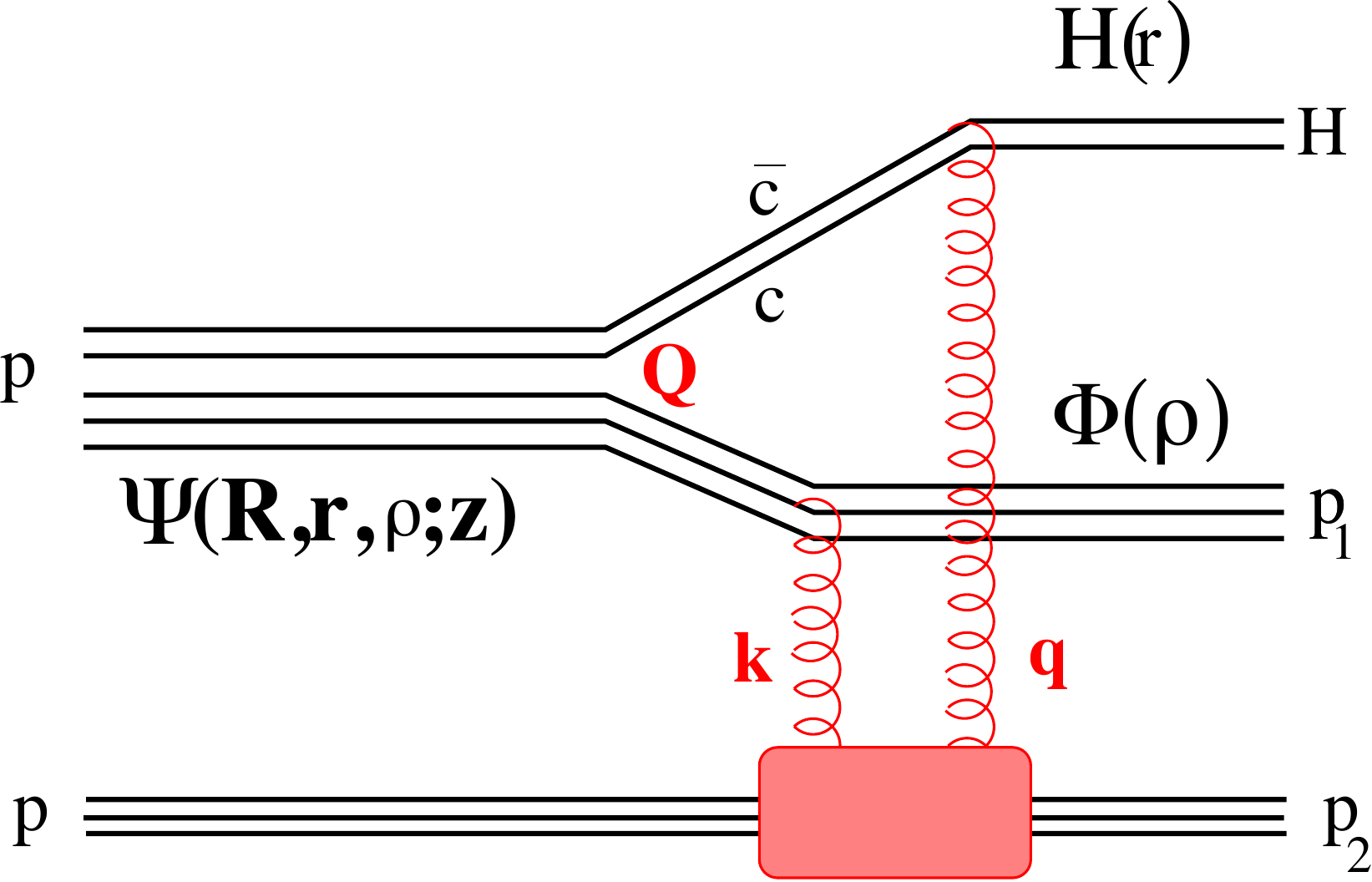}
 \includegraphics[width=5.5cm,clip]{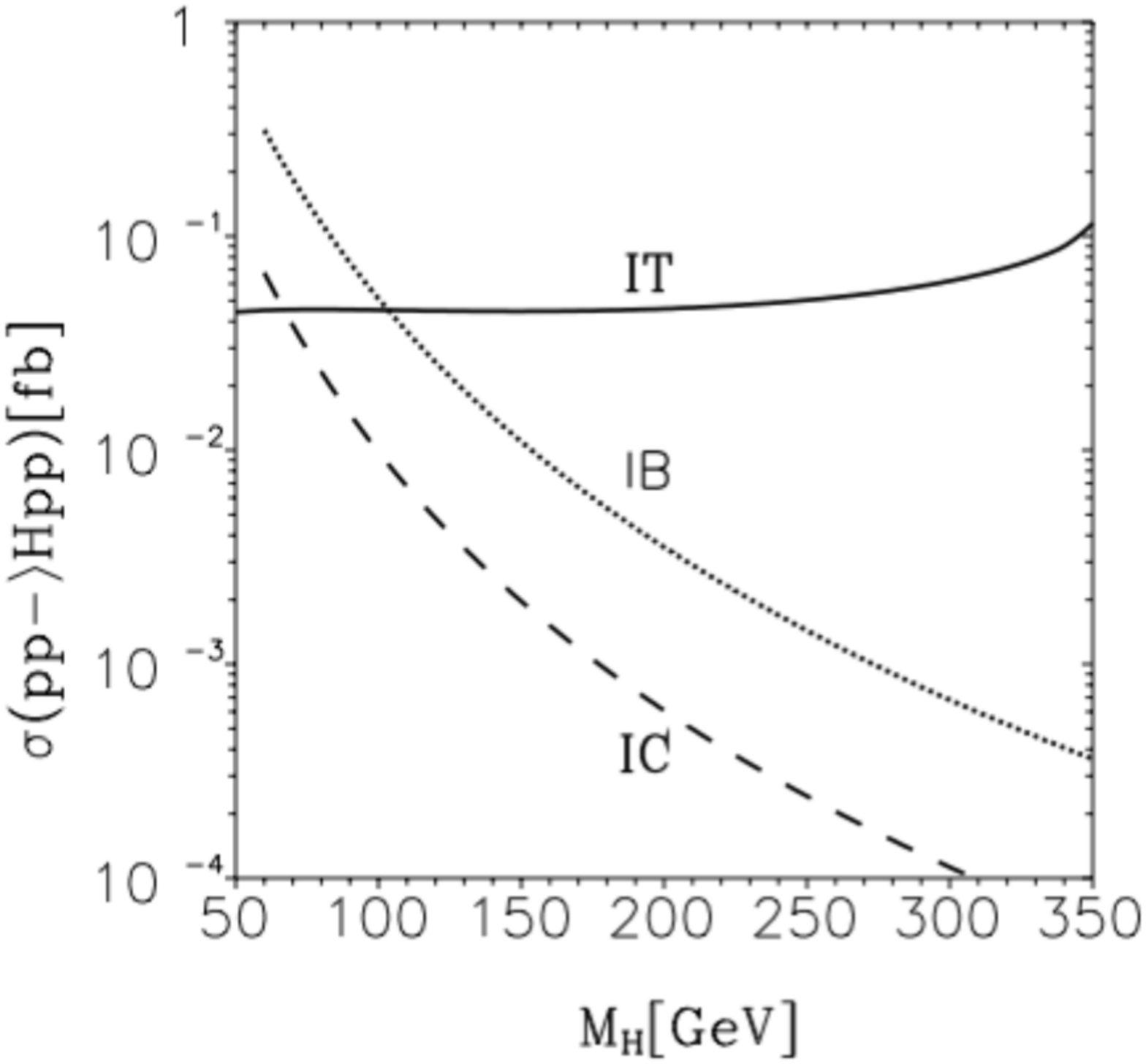}
\caption{The two-gluon exchange diagram for the Higgs exclusive production via coallicense of intrinsic heavy quarks, $\bar QQ\to H$ (left panel), and
the cross section of diffractive exclusive Higgs production off different intrinsic flavors as a function of the Higgs boson mass [\citen{brod1}] (right panel).}
\label{fig:higgs-intrinsic}      
\end{figure*}

The Higgs boson can also be diffractively produced due to fusion of the intrinsic heavy flavours (IQ) in light hadrons, $\bar QQ\to H$, 
as is depicted in Fig.~~\ref{fig:higgs-intrinsic}, left. 

Such exclusive Higgs production process, $pp\to Hpp$ was analysed in Refs.~[\refcite{brod1,brod2}].
The diffractive cross section has the form,
 \beq
 \frac{d\sigma(pp\to ppH)}
{dx_2\,d^2p_1\,d^2p_2} = \frac{1}{(1- x_2)16\pi^2}\left|A(x_2,\vec
p_1,\vec p_2)\right|^2\ ,
 \label{10}
 \eeq
 where the diffractive amplitude in Born approximation reads,
 \beqn
A(x_2,\vec p_1,\vec p_2)&=&\frac{8}{3\sqrt{2}}\, \int
d^2Q\,\frac{d^2q}{q^2}\,
\frac{d^2k}{k^2}\,\alpha_s(q^2)\alpha_s(k^2)\, \delta(\vec q +\vec
p_2 +\vec k)\, \delta(\vec k -\vec p_1 -\vec Q) \nonumber\\
&\times& \int d^2\tau\,\left|\Phi_p(\tau)\right|^2
\left[e^{i(\vec k+\vec q)
\cdot\vec\tau/2} - e^{i(\vec q-
\vec k)\cdot\vec\tau/2}\right] \int
d^2R\,d^2r\,d^2\rho\,H^\dagger(\vec r)\ e^{i\vec q\cdot\vec r/2}
\nonumber\\ &\times&
\left(1-e^{-i\vec q\cdot\vec r}\right) \Phi_p^\dagger(\vec \rho)
e^{i\vec k\cdot\vec \rho/2} \left(1-e^{-i\vec
k\cdot\vec \rho}\right)\, \Psi_p(\vec R,\vec r,\vec \rho,z)\,
e^{i\vec Q\cdot\vec R}.
 \label{15}
 \eeqn
 Here $(1-x_1)(1- x_2) = {M_H^2}/{s}$.
$\Psi_p(\vec R,\vec r,\vec \rho,z)$ is the light-cone
wave function of the IQ component of the projectile proton with transverse
separations $\vec R$ between the $\bar cc$ and $3q$ clusters, $\vec
r$ between the $c$ and $\bar c$, $\vec Q$ is the
relative transverse momentum of the $3q$ and $\bar cc$ clusters in
the projectile and $\vec\rho$ is the
transverse separation of the quark  and diquark which couple to the final-state proton $p_2.$
The density $|\Phi_p(\tau)|^2$ is the wave function of
the target proton which we also treat as a color dipole quark-diquark with transverse separation
$\tau$.  (The
extension to three quarks is straightforward [\refcite{zkl}]). The
fraction of the projectile proton light-cone momentum carried by the $\bar cc$,
$z\approx 1-x_1$. This wave function is normalized as,
 \beq
 \int\limits_0^1 dz\int d^2R\,d^2r\,d^2\rho\,
\left|\Psi_p(\vec R,\vec r,\vec \rho,z)\right|^2
=P_{IQ}\ ,
 \eeq
where $P_{IQ}$ is the weight of the IC component of the
proton, which is suppressed as $1/m_Q^2$ [\refcite{polyakov}], and is assumed to be
$P_{IC}\sim 1\%$.   The amplitudes $H(\vec r)$ and
$\Phi_p(\vec \rho)$ denote the wave functions of the produced
Higgs and the outgoing proton, respectively, in accordance
with Fig.~\ref{fig:higgs-intrinsic}, left.

At the measured Higgs mass value $125\,\GeV$ the intrinsic bottom and top provide 
comparable contributions as can be seen in Fig.~\ref{fig:higgs-intrinsic}, right. Comparing the Higgsstrahlung cross section off the produced
heavy quarks, i.e. $gg\to Q\bar Q H$, and that off the intrinsic component one concludes that the intrinsic contribution to the diffractive 
Higgs boson production can be relevant at forward Higgs boson rapidities $y_H>3.5$ [\refcite{higgs-r}].

\section{Summary}
\label{Sec:summary}

In this short review, we discussed the most important implications of the diffractive factorisation breaking in hard diffractive hadronic
collisions. Indeed, forward diffractive Abelian radiation such as radiation of direct photons, Drell-Yan dileptons, and gauge $Z^0$ and $W^\pm$ bosons 
by a projectile parton is forbidden. Nevertheless, a finite-size hadron can diffractively radiate in the forward direction due to soft interactions of 
its spectators with the target nucleon. Such a feature of the diffractive Abelian radiation breaks factorisation between soft and hard interactions 
resulting in a leading-twist behavior, i.e. to the $1/M^2$ scaling of the corresponding cross section with the boson mass $M$.

The non-Abelian forward diffractive radiation of heavy flavors is permitted even for an isolated parton. However, interaction with spectators 
provides the dominant contribution to the diffractive cross section. It comes from the interplay between large and small distances similar
to that in the diffractive Drell-Yan process. In particular, this leads to unusual properties of the SD-to-inclusive ratios such as growth with 
the hard scale $M$ and decrease with c.m. energy $\sqrt{s}$. The latter behavior holds, in fact, for various Abelian and non-Abelian radiation 
processes in diffractive hadronic collisions, in variance with predictions of diffractive factorisation. The experimental data confirm the leading-twist 
behavior of the observables.

The diffractive Higgsstrahlung off heavy flavors is possible as a double-step process when a heavy quark pair is produced first, $g \to Q\bar Q$, and then
the Higgs boson is radiated off $Q$ or $\bar Q$. Such a mechanism of diffractive Higgs boson production in association with a heavy quark pair at the LHC 
is highly suppressed ($\lesssim 1$ fb) but is not negligible compared to the conventional loop-induced central exclusive Higgs boson production. Another 
important contribution to the diffractively produced Higgs boson which dominates over the Higgsstrahlung off the produced heavy quarks comes from 
coalescence of intrinsic heavy quarks in the proton. For $M_H=125\GeV$ and forward rapidities $y_H>3.5$ dominance of intrinsic bottom and top is expected. 

Due to a decisive impact of soft spectator interactions and the universal interplay between the soft and hard interactions, the same unconventional 
hard scale and energy dependences of the SD-to-inclusive ratio has been observed in all typical diffractive reactions. This strongly motivates further
deeper phenomenological studies of diffractive factorisation breaking effects at various energies providing an access to soft dynamics of partons in a nucleon.

{\bf Acknowledgments}

This study was partially supported by Fondecyt (Chile) grants No. 1130543, 
1130549, as well as by CONICYT grant PIA ACT1406 (Chile). 
R. P. was partially supported by Swedish Research Council Grant No. 2013-4287.


\end{document}